\documentclass{IEEEtran}

\usepackage{graphicx}
\usepackage{times}
\usepackage{float}
\usepackage{eqnarray,amsmath}
\usepackage{array}
\usepackage{booktabs}
\usepackage{textcomp}
\usepackage{amsmath,amssymb,amsthm}
\usepackage{mathtools}
\usepackage{epstopdf}
\usepackage{caption}
\usepackage{multirow}
\usepackage{multicol}
\usepackage{pbox}
\usepackage{cite}
\usepackage{tikz}
\usepackage{algorithm,algpseudocode}
\usepackage{pifont}
\usepackage{dblfloatfix}
\usepackage{fixltx2e}
\usepackage{breqn}
\usepackage{xcolor}
\usepackage{subcaption}
\usepackage[font={footnotesize}]{caption}
\allowdisplaybreaks

\algnewcommand{\Inputs}[1]{%
  \State \textbf{Inputs:}\hspace*{\algorithmicindent}\parbox[t]{.8\linewidth}{\raggedright #1}
}
\algnewcommand{\Outputs}[1]{%
  \State \textbf{Outputs:}\hspace*{\algorithmicindent}\parbox[t]{.8\linewidth}{\raggedright #1}
}
\algnewcommand{\Initialize}[1]{%
  \State \textbf{Initialize:}\hspace*{\algorithmicindent}\parbox[t]{.8\linewidth}{\raggedright #1}
}

\DeclareMathOperator{\E}{\mathbb{E}}
\begin{document}
\title{A Novel Geometry-based Stochastic Double Directional Analytical Model for Millimeter Wave Outdoor NLOS Channels}
\author{Rakesh R T, \it{Student Member, IEEE},\thanks{This work was		presented in part at the IEEE International Conference on Communications, Paris, France, May 2017 \cite{rakesh2017analytical}.
		
 The authors are with G. S. Sanyal School of Telecommunications,
Indian Institute of Technology, Kharagpur, W.B -India.

e-mail: rakeshrt@gssst.iitkgp.ernet.in; debarati@gssst.iitkgp.ernet.in; gdas@gssst.iitkgp.ernet.in}~\normalfont{Debarati Sen}, \it{Senior Member, IEEE},~\normalfont{Goutam Das} \it{}}
\providecommand{\keywords}[1]{\textbf{\textit{Index terms---}} #1}
\maketitle
\begin{abstract}
	Millimeter wave (mmWave) communications which essentially employ directional antennas find applications spanning from indoor short range wireless personal area networks to outdoor cellular networks. A thorough understanding of mmWave signal propagation through the wireless channel provides valuable insights for the design of such networks which in turn dictates the achievable performance limits. High path loss, penetration loss, and
	negligible signal scattering are certain distinctive features of the mmWave channel which necessitates the development of new mathematical models. Importantly, the impact of directional antennas which spatially filter multi-path components needs to be embodied as an integral part of the channel model. In this paper, we model outdoor directional non-line-of-sight mmWave channels using a combination of stochastic geometry and image theory by expressing channel
	parameters as joint functions of transmitter-receiver separation distance, antenna half power beamwidth, and antenna beam pointing direction. By approximating the signal propagation path due to first and second order reflections from buildings, closed-form analytical expressions for average number of first order reflection components, path loss, and root-mean square delay spread of channel are derived. The accuracy of the
	model is verified by comparing numerically obtained results with experimental data reported by various urban outdoor peer-to-peer mmWave measurements, thus demonstrating the usefulness of the proposed analytical model for performance evaluation of mmWave communication networks.
\end{abstract}
\begin{keywords}
	Millimeter Wave (mmWave) Communications, Non-Line-of-Sight (NLOS), Directional Antenna, Poisson Point Process (PPP), Reflection Components.
\end{keywords}
\setlength\abovedisplayskip{0pt}
\section{Introduction}
\PARstart{T}{he} burgeoning demand for multi-Gbps data rate applications such as high definition video transmission, cloud computing, wireless backhauling, etc., has generated considerable interest in the research community to explore the under-utilized millimeter wave (mmWave) bands \cite{daniels,5grappaport,Rappa2}. Adequate bandwidth available in these frequency bands helps achieve targeted data rates even with simple modulation
schemes such as binary phase shift keying (BPSK). However, high frequency operation introduces several challenges for mmWave signal propagation. Although the reduction in link margin due to severe path loss resulting in signal propagation can be compensated by the antenna gain using directional communication \cite{Rappa2}, high penetration losses may cause links to
operate in less favorable non-line-of-sight (NLOS) conditions or even lead to outage \cite{akdeniz2014millimeter}. The challenge is particularly formidable in outdoor mmWave communication scenarios due to multiple reflections arising due to the structural layout. Therefore, the design, development, and deployment of the future generation of mmWave communication systems
entails meticulous assessment and accurate modeling of channel characteristics. Several experimental studies have been conducted to understand the behavior of directional channels in various mmWave bands for this purpose \cite{maltsev2009experimental,violette1988millimeter,kyro2012experimental,keusgen2014propagation,akdeniz2014millimeter,rappaport2015wideband}.
\vspace{-0.25cm}
\subsection{Related Work}
Measurements performed in frequency bands centered at 9.6 GHz, 28 GHz and 57 GHz \cite{violette1988millimeter} report penetration losses of the order of 100 dB which drastically affects signal propagation in these bands. Also, experiments \cite{kyro2012experimental} conducted in 81-86 GHz
band reveal that NLOS signal propagation is mainly governed by reflection components. The study also confirms that the outdoor channel in this frequency range is characterized by limited number of multi-path components. In addition, the results based on measurements conducted in the 28 GHz and
73 GHz \cite{akdeniz2014millimeter} bands indicate that the signal energy is confined in the form of spatially separated clusters, mainly generated due to distinct reflection paths. This has also been vetted by ray tracing simulations conducted for the 28 GHz \cite{hur2016proposal} and 73 GHz \cite{nguyen2014evaluation} bands which support the theory that the
signal propagation mechanism in outdoor millimeter wave channels can be accurately modeled solely based on reflection components. As for the fading phenomenon, comprehensive experimental results on large and small scale fading effects on the signal propagation in 28, 38, 60, and 73 GHz frequency
bands is available in \cite{rappaport2015wideband}. In summary, aforementioned findings confirm that mmWave channels are sparse in terms of number of multi-path components along time and angular dimensions with a significant amount of power in such components being attributed to reflections. In addition, the fact that
signal transmission through directional antennas significantly modifies multi-path characteristics also needs to be accounted for \cite{rappaport2015wideband}, \cite{rajagopal2012channel,rappaport2012cellular,rappaport201238}.

The preceding observations form the premise for development of a suitable channel model for mmWave communication. In general, channel models are broadly classified as deterministic and stochastic. Deterministic channel models use ray tracing simulations to generate channel profiles. However, since the ray tracing approach is rather intricate, stochastic channel models are more commonly used. In the stochastic channel modeling approach, the channel profile for each realization is synthesized based on the experimental data. This can be achieved in two ways: (i) parameters of each channel profile such as channel coefficients, time of arrival (ToA), angle of arrival (AoA) etc. are generated directly from probability density functions (PDF) based on the parameters obtained from experimental measurements with the aid of curve fitting techniques, (ii) channel parameters can be derived based on the knowledge of signal propagation mechanism in the wireless environment utilizing geometric statistics such as scatter distribution, scatter size etc. obtained from experimental studies. Despite their simplicity, the former modeling approaches, collectively known as non-geometric channel models \cite{rappaport2015wideband,maltsev2009experimental,akdeniz2014millimeter} pose a problem due to the need for precise setting of parameter values for specific deployment scenarios. Moreover, modeling of distance dependent as well as antenna radiation pattern dependent multi-path profile requires a large number of statistical model parameters, which may not be available in most cases. However, the latter class of channel models termed as geometric channel models have thus received favor due to the simplified control mechanism involved in generating a typical multi-path channel profile \cite{bai2014analysis,muhammad2017analytical}.

A geometry-based modeling approach is adopted in \cite{bai2014analysis} to analyze outdoor mmWave channels in which the authors evaluated the distance dependent line-of-sight (LOS) link probability by approximating buildings as rectangular objects of random size with their centers generated from Poisson
Point Process (PPP). Authors in \cite{hamalainen2007solution,li2014channel} attempted to derive the analytical expression for power delay profile (PDP) by modeling reflectors as point objects. However, these models do not capture the basic properties of reflection process due to the
point nature of the reflectors. Recently, a stochastic geometry model \cite{muhammad2017analytical} which takes only first order reflections into account characterizes mmWave PDP for omni-directional communication between the wireless nodes. Since the directional antenna is an indispensable feature in mmWave communications \cite{5grappaport,Rappa2}, the modeling of directional channel using spatial distribution of multi-path components assumes importance. The ellipse based modeling approach adopted in \cite{muhammad2017analytical} does not explicitly characterize spatial distribution of reflection components, and therefore, it is hard to obtain a tractable solution for the parameters of directional mmWave  channels. In addition, the extension of the approach to model second order reflections is also difficult. Experimental results \cite{kyro2012experimental} reveal a non-negligible impact of multi-path components due to second order reflection on received signal power as well as RMS delay spread in mmWave directional NLOS channels for most outdoor environments. In view of the aforementioned issues, a holistic modeling approach for mmWave NLOS channels is required.
\vspace{-0.3cm}
\subsection{Contribution} 
In this paper we propose a novel channel model for directional mmWave NLOS signal propagation due to first and second order reflections from building surfaces based on stochastic geometry tools \cite{haenggi2012stochastic} and image theory \cite{ballanis1997antenna}. The main objective is to explore the joint impact of antenna radiation pattern (in turn antenna half power beamwidth (HPBW)), antenna beam pointing direction, transmitter-receiver separation, building size, and building density on mmWave NLOS directional channel parameters which include the average number of reflection components, path loss, and PDP. The key contributions include:
\begin{itemize}
	\item Based on the relative positions of transmitter and receiver, and building orientation, we introduce a novel concept of feasible region that will lead to generation of first order reflection components which fall inside the antenna main lobe of both, the transmitter as well as receiver. The feasible region, determined using image theory, is useful to develop mathematically tractable models for channel parameters such as average number of reflection components, path loss, and PDP. 
	\item Since penetration losses at mmWave frequencies are reportedly high (more than $20$ dB) \cite{violette1988millimeter,bai2014analysis}, we develop models for blockage of reflection components due to buildings and human beings, and derive the corresponding expression for probability of blockage.
	\item Utilizing image theory, we extend the proposed modeling approach for second order reflections. We restrict the analysis to second order reflections because in outdoor mmWave channels, it has been observed that reflections beyond second order do not contribute much to the received signal power  \cite{kyro2012experimental,rappaport2015wideband}.
	\item Based on the proposed model for reflection, we derive closed form expressions for average number of first order reflections, path loss, PDP of directional mmWave channels and thereby find the average multi-path delay, RMS delay spread, and coherence bandwidth of the channel.
\end{itemize}
The rest of the paper is organized as follows: A detailed system model with various assumptions considered in the paper is provided in Section II. In Section III, we present statistical characterization of first order reflections under directional NLOS signal propagation. Modeling of second order reflection process and the derivation of PDP based on first and second order reflection processes are given in Section IV. A comparison of analytical results with simulated as well as experimental results is elaborated in Section V. Finally, Section VI concludes the paper.
\section{System Model}
\begin{figure}[H]\label{Deploy_scenario}
	\centering
	\includegraphics[trim=6.8cm 7cm 8.2cm 5.1cm, clip=true, totalheight=0.21\textheight]{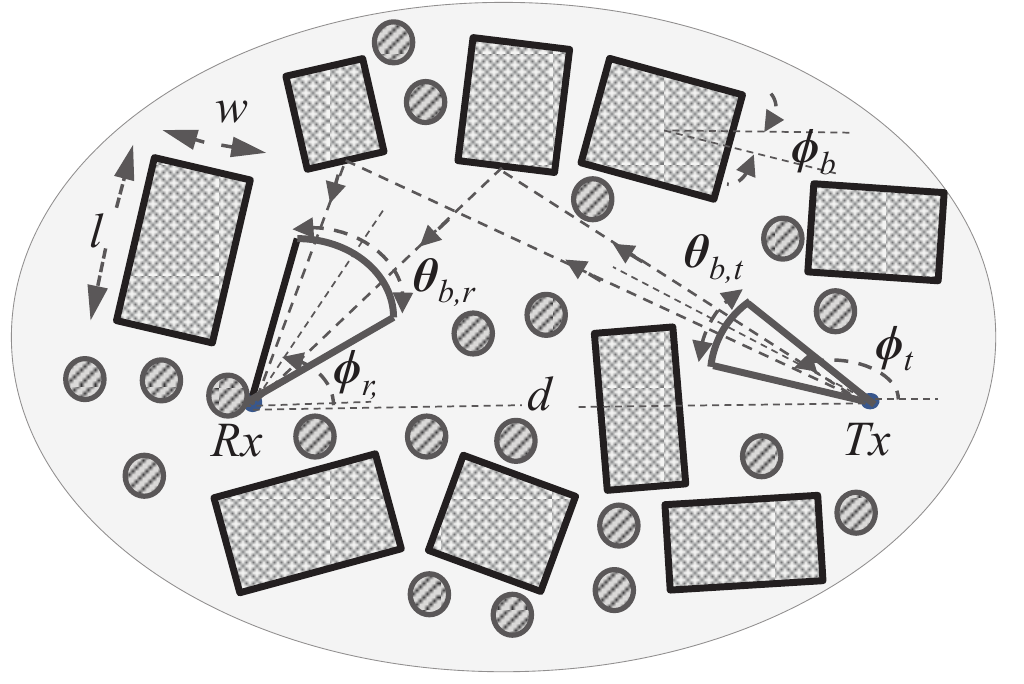}
	\caption{Deployment scenario}	
	\label{fig:Deployment scenario}
\end{figure}

We consider an outdoor scenario where transmitter and receiver nodes are deployed in a 2D plane as shown in Fig.~\ref{fig:Deployment scenario}. The Euclidean distance between the transmitter and the receiver is denoted by $d$. We assume that all nodes are equipped with directional antennas. For analytical tractability, the antenna radiation pattern is approximated using the cone model with constant main lobe gain $G_{m,i}=\frac{2\pi}{\theta_{b,i}}$ and zero side lobe gain \cite{bai2015coverage}, where $\theta_{b,i}$ denotes antenna HPBW with $i=r$ and $i=t$ for receiver and transmitter nodes, respectively.
The error introduced by the cone model can be minimized by adjusting  main lobe gain and HPBW of the cone model with respect to the radiation pattern of a given directional antenna. The antenna beam pointing direction for transmitter and receiver are represented by angle terms $\phi_{t}$ and $\phi_{r}$, respectively, which are denote the angle made by the main lobe axis with the horizontal line joining the points $Tx$ and $Rx$ as depicted in Fig.~\ref{fig:Deployment scenario}.
The nodes are surrounded by buildings with varying size and orientation. The transmitter-receiver nodes are assumed to engage in device-to-device communication or constitute part of a short range infrastructure based communication network. For such scenarios, the height of transmitter as well as receiver antenna from the ground surface is very less compared to the typical building height. Hence,
we model the buildings as rectangles with length $l$ and width $w$ \cite{bai2014analysis}. The length and width of all buildings follow identical and independent distributions, $f_{L}(l)$ and $f_{W}(w)$, respectively. The building orientation is modeled by the uniformly distributed angular parameter $\phi_{b}$ with $f_{\Phi_{b}}(\phi_{b})\sim U(0,\pi )$, and the center of each building is generated from a PPP \footnote[1]{It is interesting to note that for larger building densities, the PPP assumption may cause overlap of building geometry thereby introducing error in modeling. However, the assumption is viable since the results for channel parameters using this model indicate a fairly good match with experimental data obtained in a typical urban environment \cite{rappaport201238}.} with density $\lambda_{b}$.
The signal propagation occurs either through LOS or in the event of blockage, through NLOS. In this work, we consider signal propagation through NLOS and the dominant mode which may either occur due to first or second order reflection from the smooth building surface \cite{kyro2012experimental,rappaport2015wideband}. Modeling of objects such as lamp posts, vehicles, etc. have not been considered in our work since it has been established that dominant reflections from smooth building surfaces alone is good enough to characterize typical urban outdoor NLOS environments \cite{hur2016proposal}. It should be noted that reflections from building surfaces may in turn be blocked by other objects including buildings\footnote[2]{In a typical urban environment, reflection paths may be blocked by vehicles as well. The geometry of vehicles can be modeled as rectangular objects in a manner similar to the model for buildings.}, human beings etc. In this paper, we do model this blockage event.

As depicted in Fig.~\ref{fig:Deployment scenario}, the circular discs of diameter $W_{h}$ represent a human being with center point of each such disc, i.e., their respective location of each person, realized through a PPP of density $\lambda_{h}$.\footnote[3]{We note that human beings present inside buildings do not block outdoor reflection components. Similar to the approach considered in \cite{bai2015coverage}, we assume that the  spacial distribution of human beings outside the buildings is a thinned PPP with density  $\lambda_{h}=\lambda_{h}^{'}(1-\lambda_{b}\E[l]\E[w])$, where $\lambda_{h}^{'}$ denotes density of human beings before thinning being applied.}
The height of human beings is a significant factor in the occurrence of blockage events. At a considerable distance from the location of base station, however, a human being may not block a reflection path due to the fact that the reflection component will pass above the person.
Nevertheless, the height of a human being can be incorporated into the analysis by applying independent non-uniform thinning of PPP which governs spatial distribution of human beings using an approach similar to that adopted in \cite{bai2014analysis}. The statistical model for generation and blockage of first order reflections is described in the following section.
\section{Characterization of first order reflections in directional NLOS transmission}
In this section, we present a reflection model based on image theory to characterize first order reflections. We introduce the notion of feasible region for transmitter and receiver antenna main lobe coupling, and derive the expression for the  area of this region. The probability of blockage of a reflected path is then evaluated following which, the statistical characterization for number of first order reflections along with the expression for overall path loss for directional NLOS transmission.
\subsection{Geometric modeling of first order reflections in mmWave directional channels}
\begin{figure}[H]\label{sys_model}
	\centering
	\includegraphics[trim=1cm 2.0cm 2.6cm 1.2cm, clip=true, totalheight=0.22\textheight]{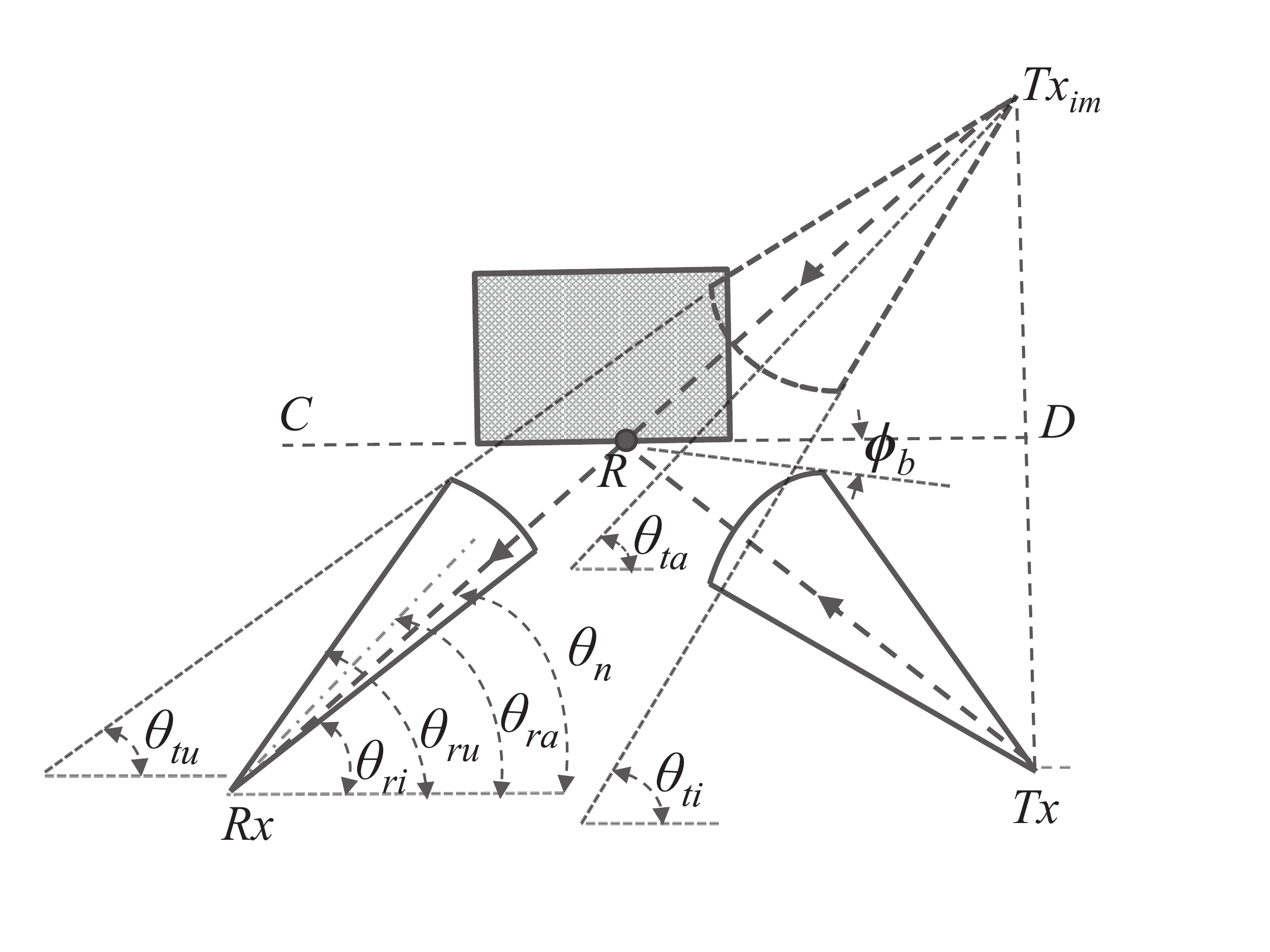}
	\caption{Reflection model based on image theory}	
	\label{fig:System Model}
\end{figure} 
To model first order reflections, we develop a novel approach based on image theory as illustrated in Fig.~\ref{fig:System Model}. The reflector (building) with smooth surface is located in the 2D plane with orientation angle $\phi_{b}$. To aid better visualization for the ensuing mathematical analysis, we rotate the entire topology by an angle $\phi_{b}$ such that the reflecting surface is aligned with the horizontal axis (see Fig.~\ref{fig:System Model}). In the proposed model, the image of the transmitter node (${Tx_{im}}$) is generated as its reflected duplication visible in the direction perpendicular to the reflector surface.
Therefore, ${Tx_{im}}$ is located in the 2D plane in such a way that the line segment joining the transmitter point and image point, $Tx_{im}-Tx$ is perpendicular to the line segment $CD$ with the midpoint of $Tx_{im}-Tx$ located along this line segment itself. 
Based on this model, it is clear that a valid reflection results only when the line segment $Rx-Tx_{im}$ intersects with the reflecting surface (at point $R$). The reflected path located within the transmitter and receiver antenna main lobes is taken into consideration and rest of the paths are discarded. In Section III-B, we show the existence of a feasible region; if the center of a building is located inside this region, it results in a reflected path due to antenna main lobe coupling. The angular parameters are obtained from the right angle triangles
in Fig.~\ref{fig:System Model} as $\theta_{ri}=\phi_{r}+\phi_{b}-\theta_{b,r}/2$, $\theta_{ru}=\phi_{r}+\phi_{b}+\theta_{b,r}/2$,
$\theta_{ti}=\pi-\left(\phi_{t}+\phi_{b} -\theta_{b,t}/2\right)$, $\theta_{tu}=\pi-\left(\phi_{t}+\phi_{b}+\theta_{b,t}/2\right)$,
$\theta_{ra}=\phi_{r}+\phi_{b}$, $\theta_{ta}=\pi -\left(\phi_{t}+\phi_{b}\right)$. $\theta_{n}$ represents the AoA of the reflected path at the receiver.
\subsection{Evaluation of feasible area for main lobe coupling}
In this section, we calculate the area of the feasible region which depends on the main lobe coupling between the transmitter and receiver nodes for a specified separation distance, $d$, antenna beam pointing angles, $\phi_{r}$ and $\phi_{t}$, building orientation angle, $\phi_{b}$, and HPBW, $\theta_{b}$.
The evaluation of the feasible region area enables the derivation of average number of reflections, path loss, and PDP for directional transmission. \newline \textbf{\textit{Theorem 1}:} Subject to fixed values of transmitter-receiver separation $d$, building dimensions $w$ and $l$, antenna pointing angles $\phi_{t}$, $\phi_{r}$, $\phi_{b}$, and antenna HPBW $\theta_{b}$, area of the feasible region is given by,\\
\begin{align}\label{prob_coup}
A=a\frac{d|\text{cos}\phi_b|}{2}\left(\text{tan}\theta_{u}
-\text{tan}\theta_{i}\right)
\end{align}
where $a=l$, if the reflection occurs along the length of the building and $a=w$, if the reflection occurs along its width. $\theta_{i}$ and $\theta_{u}$ denote the lower and upper limits for $\theta_{n}$, respectively. Exact values for $\theta_{i}$ and $\theta_{u}$ are determined at the end of this section.
\begin{figure}[H]\label{Analytic1}
	\centering
	\includegraphics[trim=1cm 0.9cm 1.1cm 1.6cm, clip=true, totalheight=0.23\textheight]{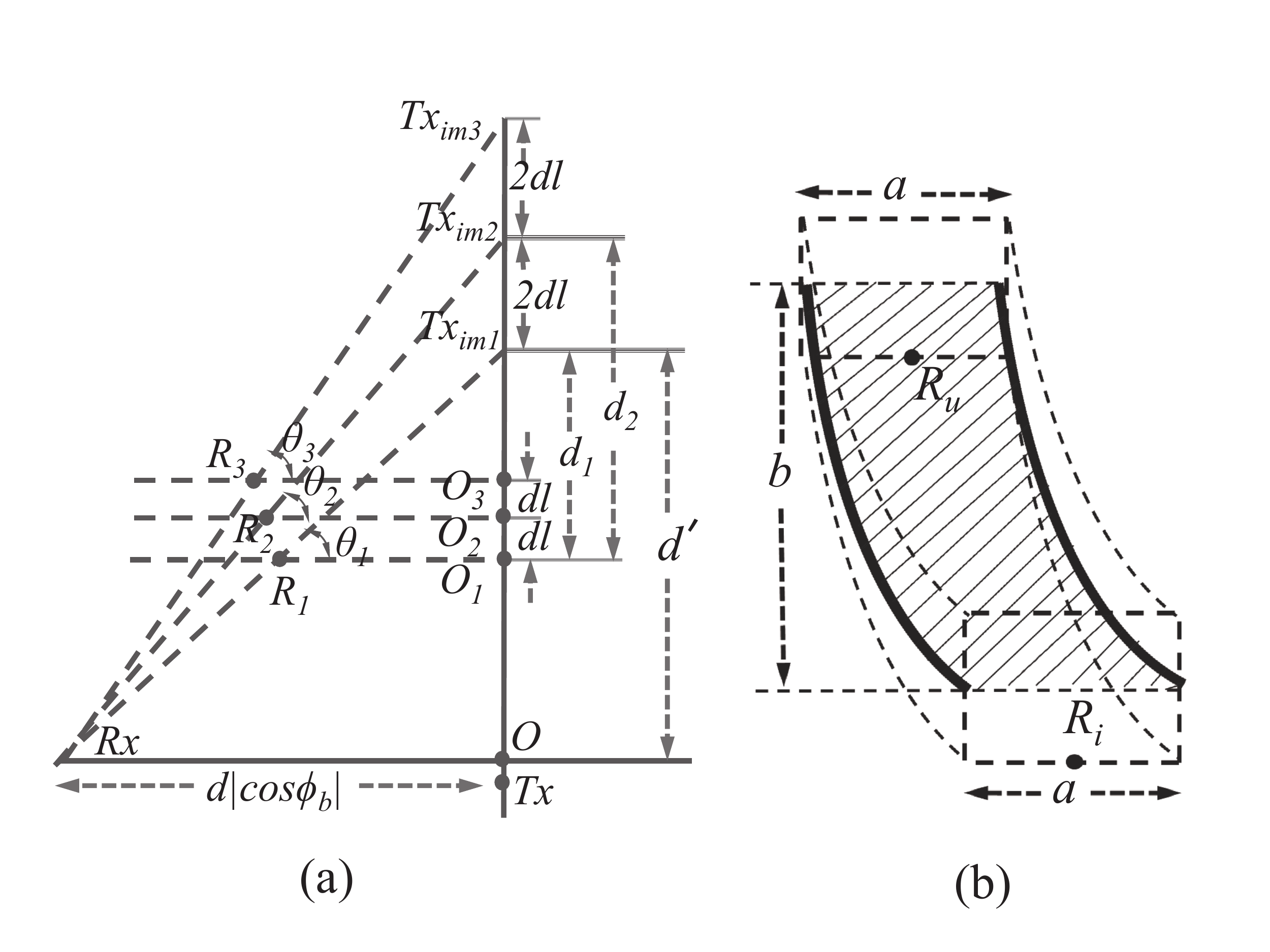}
	\caption{Geometrical interpretation of the reflection process: (a) Movement of the reflection points for a fixed $\phi_{b}$
		(b) Feasible region for main lobe coupling}	
	\label{fig:Analytic1}
\end{figure}
\begin{proof}
\begin{figure*}[!b]
\hrule
\setcounter{equation}{4}
\begin{align}\label{tot_area}
A_{block}=\text{Area}(ABWXFYPDE)=&\text{Area}(ABCD)+\text{Area}(WXYZ)+\text{Area}(AED)+\text{Area}(XYF)
+\text{Area}(BRW)\nonumber\\&-\text{Area}(ZRC)-\text{Area}(ZPC).
\end{align}
\setcounter{equation}{1}
\end{figure*}
To identify the feasible region for a given orientation angle $\phi_{b}$, we model the position of a reflector (in this case, a building) in terms of its center point along the vertical or horizontal direction (as shown in Fig.~\ref{fig:System Model}). We observe that the point of reflection $R$ remains unchanged if the position of the reflector is translated in horizontal direction (this is due to the law of reflection).  However, any change in position of the reflector along the vertical direction creates a trajectory of reflection points $R_{1}$, $R_{2}$, ..., $R_{n}$ as illustrated in Fig.~\ref{fig:Analytic1}(a).
In order for a reflector to generate a reflection that leads to transmitter-receiver antenna main lobe coupling, we note that there exists limits for its possible positions along each direction. In the horizontal direction, if the distance separation between the point of reflection $R_{i}$ and the edge point of the reflector is greater than $a/2$, then the reflected path does not lie within the main lobe of the receiver. We see that such lower and upper limits for reflector positions also exist along the vertical direction within which main lobe coupling occurs.
To evaluate the area of feasible region for main lobe coupling, we first determine the trajectory of the points $R_{1}$, $R_{2}$, ..., $R_{n}$, if the location of the reflector is translated by an incremental amount $dl$ as shown in Fig.~\ref{fig:Analytic1}(a). The distance $d_{n}$ for the $n$-th shift from location $R_{1}$ is,\\
\begin{align}
d_{n}=\frac{d|\text{cos}\phi_{b}|\left(d_{1}+ndl\right)}{\left(d'+2ndl\right)}.
\end{align}
The feasible region for antenna main lobe coupling is highlighted by the shaded area in Fig.~\ref{fig:Analytic1}(b) with end points for the change in the position of the reflector along the vertical direction being determined by its $i$-th and $u$-th shift values. The corresponding points of reflection are denoted by $R_{i}$ and $R_{u}$.
The shaded region in Fig.~\ref{fig:Analytic1}(b) can be mapped into an equivalent rectangular region with area $A=ab$ due to the parallel edges of the feasible region. The length $b$, which is the same as the vertical distance between reflection points $R_{i}$ and $R_{u}$, can be evaluated from Fig.~\ref{fig:Analytic1}(a) as,\\
\begin{align}
b&=\left[\left(d'+2udl\right)-\left(d_{1}+udl\right)\right]-\left[\left(d'+2idl\right)-\left(d_{1}+idl\right)\right]\nonumber\\
&=\left[d|\text{cos}\phi_{b}| \text{tan}\theta_{u}-\left(d|\text{cos}\phi_{b}|/2\right)\text{tan}\theta_{u}-
\left(OTx/2\right)\right]\nonumber\\&\hspace{0.5cm}-\left[d|\text{cos}\phi_{b}|
\text{tan}\theta_{i}-\left(d|\text{cos}\phi_{b}|/2\right)\text{tan}\theta_{i}-\left(OTx/2\right)\right]\nonumber\\
&=\left(d|\text{cos}\phi_{b}|/2\right)\left(\text{tan}\theta_{u}-\text{tan}\theta_{i}\right),
\end{align}
where $OTx$ is the distance between the points $O$ and $Tx$. We note that the points $Tx_{im}$ and $Tx$ are equidistant from the reflecting plane. Hence, area of the feasible region is given by,
$A=a\frac{d|\text{cos}\phi_b|}{2}\left(\text{tan}\theta_{u}-\text{tan}\theta_{i}\right)$
\end{proof}
It should be noted that lower and upper limits $\theta_{i}$ and $\theta_{u}$ are related to the modified transmitter and receiver antenna beam pointing angle $\phi_{ta}$ and $\phi_{ra}$, respectively. They are evaluated as $\theta_{i}=\text{max}(\theta_{ri},\theta_{ti})$ and $\theta_{u}=\text{min}(\theta_{ru},\theta_{tu})$. For $\theta_{b,t}=\theta_{b,r}=\theta_{b}$, the limits for $\theta_{i}$ and $\theta_{u}$ can be found by considering two scenarios, i.e. $\theta_{ta}>\theta_{ra}$ (Scenario 1) and  $\theta_{ra}>\theta_{ta}$ (Scenario 2). For Scenario 1, we evaluate $\theta_{i}=\theta_{tu}$ and $\theta_{u}=\theta_{ru}$, and the region outside this range does not lead to main lobe coupling. The same is true for Scenario 2 with $\theta_{i}=\theta_{ri}$ and $\theta_{u}=\theta_{ti}$.
We note that the condition $\theta_{ri}\geq\theta_{ti}$ or $\theta_{ru}\leq\theta_{tu}$ also does not lead main lobe coupling.

Based on the preceding analysis, we can statistically characterize the number of first order reflections. Due to law of reflection, it is clear that every smooth reflector surface may generate a single first order reflection due to antenna main lobe coupling. Therefore, number of reflections captured by the antenna main lobe of the receiver is same as the number of reflectors which yield feasible reflections.
As such, the number of reflections follows a Poisson probability density function (PDF) with parameter $\lambda_{b}A$ (with fixed $\phi_{b}$). However, certain reflections may undergo blockage, and thereby statistical parameters of the reflections may change, which is discussed in Section III-C and Section III-D.
\subsection{Evaluation of probability of blockage of a reflecting path}
Since a reflected path may be blocked by buildings and human beings independently, we evaluate corresponding probability of blockage separately in the following subsections.
\subsubsection{Probability of blockage of a reflecting path due to buildings}
The following theorem gives the blocking probability for each reflection path.
\newline \textbf{\textit{Theorem 2}:} The blocking probability for a given reflection due to buildings conditioned on $d$, $\phi_{b}$, and $\theta_{n}$ is given by,\\
\begin{align}
P_{b,B/\phi_{b},\theta_{n}}=1-\text{exp}(-\lambda_{b}A_{block}),
\end{align}
where $A_{block}\approx d\left|\text{cos}(\phi_{b})\right|\text{sec}\theta_{n}\frac{2}{\pi}\left(\E\left[l\right]
+\E\left[w\right]\right)+\E[w]\E[l]-\frac{\left(2-\left(\text{cos}\theta_{n}+\text{sin}\theta_{n}\right)\right)}
{2\pi}\left(\E\left[w^{2}\right]+\E\left[l^{2}\right]\right)$.
\setcounter{equation}{12}
\begin{figure}[H]\label{block_prob}
	\centering
	\includegraphics[trim=3.4cm 2.4cm 2.2cm 2.4cm, clip=true, totalheight=0.24\textheight]{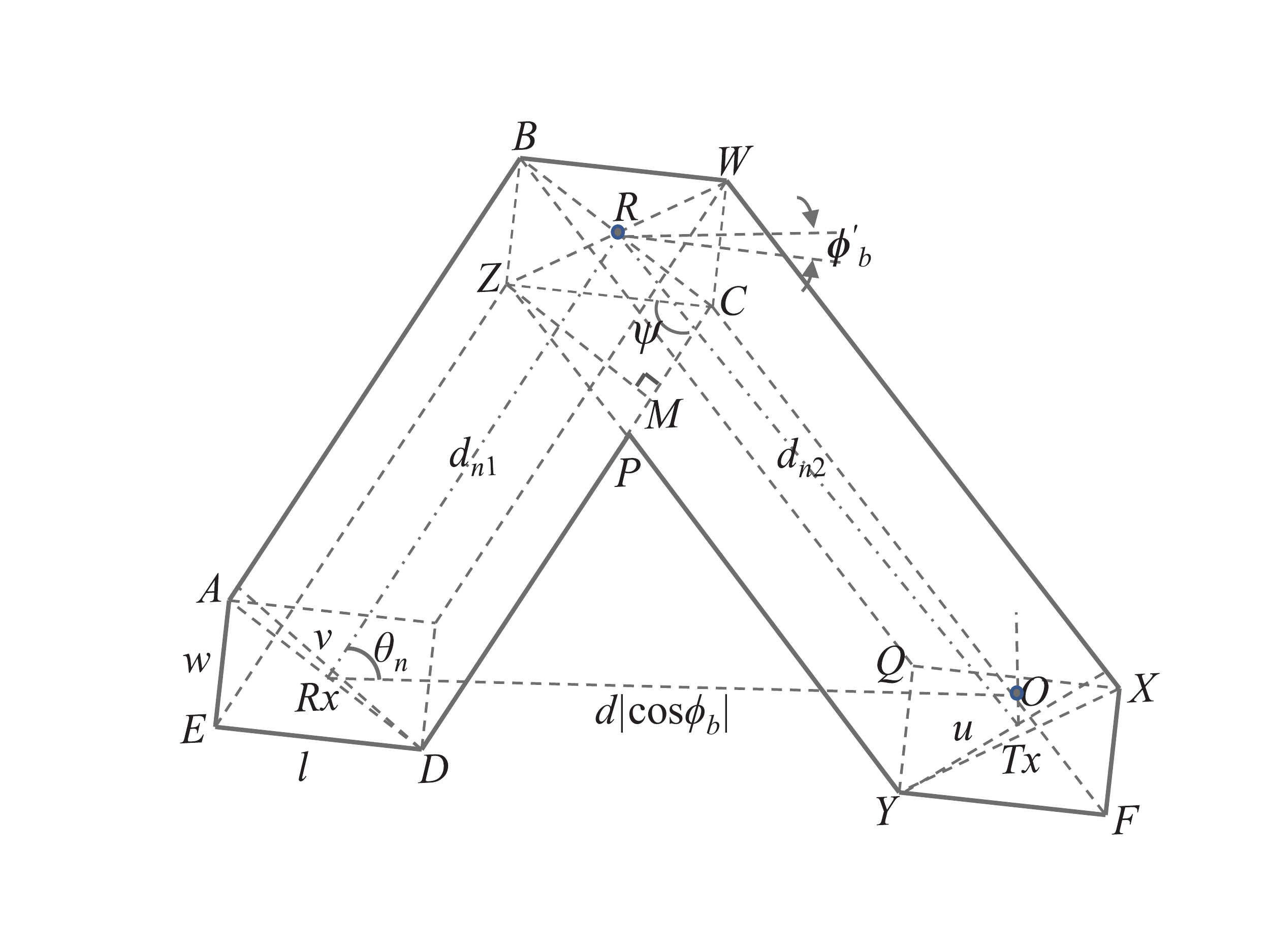}
	\caption{Blockage scenario for a typical reflected path due to buildings.}	
	\label{fig:Blockage scenario}
\end{figure}
\begin{proof}
The effect of blockage of reflected paths due to buildings is evaluated based on the model provided in Fig.~\ref{fig:Blockage scenario}, which is an approach similar to the one provided in \cite{muhammad2017analytical}. The evaluation of blocking probability in this paper is applicable general irrespective of the building orientation angle. We note that the model adopted in \cite{muhammad2017analytical} is a special case (which assumes same orientation angle for all buildings) of the model considered in this paper.
We calculate the probability of blockage conditioned on the orientation angles for reflector and blockage ($\phi_{b}$ and $\phi_{b,B,block}$, respectively). We further construct a geometry around the reflected path $TxRRx$ as shown in Fig.~\ref{fig:Blockage scenario}. If the center point of a building is located inside this bounded area, the reflected path is blocked. Since the reflector points are distributed according to PPP, blocking probability is calculated from the area of the geometry using the null probability of Poisson distribution.
It should be also noted that the geometry in Fig.~\ref{fig:Blockage scenario} is rotated through the angle $\phi_{b}$ similar to that in Fig.~\ref{fig:System Model} with $\phi_{b'}=\phi_{b}-\phi_{b,block}$, and conditioned on $\phi_{b}$, $\phi_{b'}\sim U(-\pi,\pi)$. Overall area of the geometry $A_{block}$ is expressed as (\ref{tot_area}). The derivation of individual components in $A_{block}$ is available in the conference version of our paper \cite{rakesh2017analytical}, and the result reproduced as: Area($ABCD$)+Area($WXYZ$)$=\frac{2}{\pi}d|\text{cos}\phi_{b}|\text{sec}\theta_{n}[\E(l)+\E(w)]$, Area($AED$)$+$Area($XYZ$)$=\E(l)\E(w)$, Area($BRW$)$=$Area($ZRC$), and Area($ZPC$)$\approx \frac{(2-(\text{cos}\theta_{n}+\text{sin}\theta_{n}))}{2\pi}[\E(w^{2})+\E(l^{2})]$. Therefore, probability of blockage for a first order reflection, $P_{b,B/\phi_{b},\theta_{n}}$, due to buildings, determined using null probability of PPP can be expressed as, $P_{b,B/\phi_{b},\theta_{n}}=1-\text{exp}(-\lambda_{b}A_{block})$\end{proof}
\begin{figure*}[!b]
	\setcounter{equation}{7}
	\hrule
	\vspace{0.2cm}
	\begin{align}\label{N_r4}
	N_{r}=\frac{1}{\pi}\left[\left(\int_{\phi_{i1}}
	^{\phi_{u1}}N_{r,1/\phi_{b}}\text{d}\phi_{b}
	+\int_{\phi_{i2}}
	^{\phi_{u2}}N_{r,2/\phi_{b}}\text{d}
	\phi_{b}\right)_{a=l}\hspace{-0.25cm}+\left(\int_{\phi_{i1}}
	^{\phi_{u1}}N_{r,1/\phi_{b}}\text{d}\phi_{b}
	+\int_{\phi_{i2}}
	^{\phi_{u2}}N_{r,2/\phi_{b}}\text{d}
	\phi_{b}\right)_{a=w}\right]
	\end{align}
	\setcounter{equation}{10}
	\begin{align}\label{N_r3}
	N_{r/\phi_{b}}=&\frac{\lambda_{b}d(P_{self})^{i}}{2\left|\text{cos}\phi_{b}\right|^{-1}}\E_{L,W}\bigg[ a\bigg(\bigg(1+\bigg(x+\frac{2u_{0}z}{\sqrt{1+u_{0}^{2}}}\bigg)u_{0}\bigg)(\text{tan}\theta_{u}-\text{tan}\theta_{i})-\bigg(\frac{x}{2}+\frac{u_{0}z}{\sqrt{1+u_{0}^{2}}}\bigg)\bigg[\text{tan}^{2}\theta_{u}-\text{tan}^{2}\theta_{i}\bigg]\bigg)\nonumber\\&\hspace{12cm}\times e^{-xu_{0}-y-z\sqrt{1+u_{0}^{2}}}\bigg].
	\end{align}
\end{figure*}
\begin{figure}[H]
	\centering
	\includegraphics[trim=3.8cm 2.9cm 2.2cm 3.5cm, clip=true, totalheight=0.21\textheight]{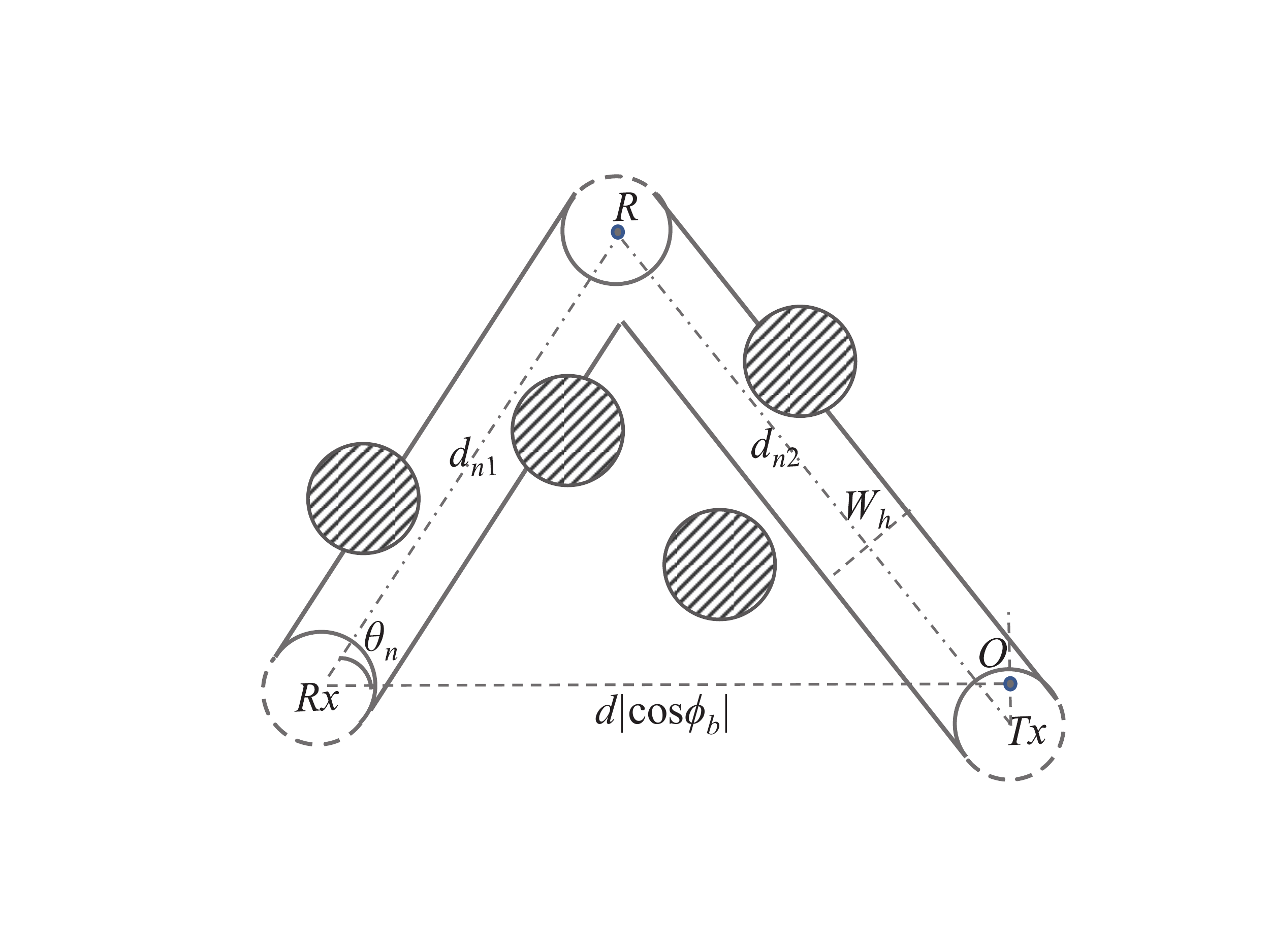}
	\caption{Blockage scenario for a typical reflected path due to human beings.}	
	\label{fig:Blockage scenario_humans}
\end{figure}
\vspace{-0.25cm}
\subsubsection{Probability of human blockage of a reflecting path}\footnote[4]{Modeling of human blockage for reflection components in enclosed environments can be found in \cite{george2017enclosed}. However, this modeling approach cannot be directly apply to outdoor scenarios.}
Similar to the model for blockage of reflection path by buildings, we construct a blocking area based on a shape approximation around the reflection path as shown in Fig.~\ref{fig:Blockage scenario_humans}. Unlike blockage due to buildings, the overlapping area near the point of reflection $R$ and the exclusion regions around the transmitter-receiver nodes are negligible as compared to total blocking area (this is due to the large propagation distance of the reflection component). Therefore, we approximate the blocking region in terms of a rectangle with length $d|\text{cos}\phi_{b}|\text{sec}\theta_{n}$ (corresponding to length of the reflected path) and width $W_{b}$.
Hence, the probability of blockage of reflection path in this case is given by $P_{b,H/\theta_{n}}=1-\text{exp}(-\lambda_{h}W_{h}d|\text{cos}\phi_{b}|\text{sec}\theta_{n})$. We note that the user holding the wireless node may also block the reflection path resulting in a \textit{self-blockage} event, which can be modeled by a constant non-blocking probability term $P_{self}$ as obtained in \cite{george2017enclosed}.
Since blockage due to buildings and human beings are independent events, overall blocking probability of first order reflection path is derived as,\\
\setcounter{equation}{5}
\begin{align}
P_{b/\phi_{b},\theta_{n}}=&1-(P_{self})^{i}\left(1- P_{b,B/\phi_{b},\theta_{n}}\right) \left(1-P_{b,H/\theta_{n}}\right) \nonumber\\
=&1-(P_{self})^{i}\text{exp}(-\lambda_{b}A_{block}-\lambda_{h}W_{h}d|\text{cos}\phi_{b}|\text{sec}\theta_{n})
\end{align}
where $i=1$, if either the transmitter node or receiver node is carried by a person, $i=2$, if both nodes are carried by respective individuals, and $i=0$, if the nodes are placed in some fixed location.
\subsection{Statistical characterization of number of first order reflections}
Thus far, we have shown that the number of first order reflections under zero blocking probability follows Poisson distribution with average number of first order reflections equal to $\lambda_{b}A$, where $A$ denotes the feasible area for antenna main lobe coupling. In this section, we analyze the statistical nature of first order reflections under blockage events. The blockage of reflection components results in distance and angle dependent thinning of points in the feasible area $A$ (see the expression for $P_{b/\phi_{b},\theta_{n}}$).
Hence, we approximate the resulting thinned point process for feasible reflections as an inhomogeneous PPP with intensity $\lambda_{b}\left(1-P_{b/\phi_{b},\theta_{n}}\right)$ \cite{haenggi2012stochastic}[ch. 2.7.3]. In fact, the same PPP generates the center points of reflector as well as blockages. Apart from the human blockage, existence of a feasible reflection path is governed by the mutual coupling between points of the parent PPP which models the spatial distribution of buildings. As a result, the thinned point process corresponding to the feasible reflections cannot be modeled as a PPP.
However, it should be noted that if transmitter-receiver separation, $d$, is sufficiently large, then the area $A$ is much smaller as compared to $A_{block}$. Therefore, the spatial distribution of locations of reflectors which generate feasible reflections can still be approximated as PPP with a modified density parameter. To support the derivation for average number of reflections for a given transmitter-receiver separation $d$, antenna HPBW $\theta_{b}$, transmitter-receiver antenna beam pointing angles $\phi_{t}$ and $\phi_{r}$, we first propose following lemmas.\vspace{0.25cm}
\newline \textbf{\textit{Lemma 1}:} Let $\Phi(l,w,\phi_{b})$ be a PPP formed by centers of
the buildings with dimensions in ($l,l+\text{d}l$), ($w,w+\text{d}w$), and orientation angles in ($\phi_{b},\phi_{b}+\text{d}\phi_{b}$). Then, $\Phi(l,w,\phi_{b})$ is a PPP with density $\lambda_{l,w,\phi_{b}}=\lambda_{b}f_{L}(l)\text{d}lf_{W}(w)\text{d}wf_{\Phi_{b}}(\phi_{b})\text{d}\phi_{b}$. If $\lambda_{l_{1},w_{1},\phi_{b1}}\neq\lambda_{l_{2},w_{2},\phi_{b2}}$, then $\Phi(\lambda_{l_{1},w_{1},\phi_{b1}})$, $\Phi(\lambda_{l_{2},w_{2},\phi_{b2}})$ are independent PPPs (The proof is similar to Lemma 1 provided in \cite{bai2014analysis}). 

\hspace{-0.4cm}\textbf{\textit{Lemma 2}:} Let $K_{r}(l,w,\phi_{b})$ denote the number of unblocked reflections generated by the building PPP $\Phi(l,w,\phi_{b})$. Then, $K_{r}(l,w,\phi_{b})$ is approximately a Poisson random variable with parameter $N_{r}(l,w,\phi_{b})=\lambda_{b}\int_{A}\left(1-P_{b/\phi_{b},\theta_{n}}\right)\text{d}A$, where $A$ is a function of $l$, $w$, and $\phi_{b}$.
\begin{proof}
 Based on the assumption, $A<<A_{block}$, $K_{r}(l,w,\phi_{b})$ can be approximated as a Poisson random variable with average number of feasible reflections $N_{r}(l,w,\phi_{b})=\lambda_{b}\int_{A}\left(\left(1-P_{b/\phi_{b},\theta_{n}}\right)\right)\text{d}A$ (the independence of the two PPPs - i.e. due to human beings and buildings, helps retain $K_{r}(l,w,\phi_{b})$ as a Poisson random variable). The elemental area corresponding to the feasible area $A$ is denoted by $\text{d}A$ (derivation of $\text{d}A$ is provided in Appendix A). 
\end{proof}
We now derive the expression for average number of reflections using the following theorem.\vspace{0.25cm}
\newline \textbf{\textit{Theorem 3}:} Let $K$ denote the number of first order reflections within the antenna main lobe of a receiver subject to transmitter-receiver separation $d$, antenna HPBW $\theta_{b}$, transmitter-receiver antenna beam pointing angles $\phi_{t}$ and $\phi_{r}$. Then, $K$ can be modeled as a Poisson random variable with probability mass function given by, $P_{r}\left(K\right)=\frac{\left(N_{r}\right)^{K}\exp(-N_{r})}{K!},$
where $N_{r}=\frac{\lambda_{b}}{\pi}\int_{\phi_{b}}\E_{L,W}\left[\int_{A}\left(1-P_{b/\phi_{b},\theta_{n}}\right)\text{d}A\right]$$\text{d}\phi_{b}
$ represents the average number of first order reflections. $\E_{L,W}\left[.\right]$ denotes expectation over the distributions $f_{L}(l)$ and $f_{W}(w)$.
\begin{proof}
Since superposition of Poisson random variables is also a Poisson random variable, the average number of valid reflections $N_{r}$ can be determined as,\\
\begin{align}\label{no_ref}
N_{r}=&\sum_{l,w,\phi_{b}}N_{r}(l,w,\phi_{b})=\int_{\phi_{b}}\E_{L,W}\left[N_{r}(l,w,\phi_{b})\right]\frac{1}{\pi}\text{d}\phi_{b}\nonumber\\
=&\frac{\lambda_{b}}{\pi}\int_{\phi_{b}}\E_{L,W}\left[\int_{A}\left(1-P_{b/\phi_{b},\theta_{n}}\right)\text{d}A\right]\text{d}\phi_{b}.
\end{align}
\end{proof}
The average number first order reflections is determined by substituting the expression of $P_{b}/\phi_{b},\theta_{n}$ and $\text{d}A$ from Appendix A in (\ref{no_ref}), which results in (\ref{N_r4}) (we assume $\theta_{b,r}=\theta_{b,t}=\theta_{b}$). $N_{r,1/\phi_{b}}$ and $N_{r,2/\phi_{b}}$ denote the average number of first order reflection components corresponding to two scenarios described in Section III-B), respectively. The limits for $\phi_{b}$ are evaluated based on the conditions mentioned in Section III-B.
i.e., $\phi_{i1}=\left(\pi-\phi_{t}-\phi_{r}-\theta_{b}\right)/2$, $\phi_{u1}=\left(\pi-\phi_{t}-\phi_{r}\right)/2$,  $\phi_{i2}=\phi_{u1}$, and $\phi_{u2}=\left(\pi-\phi_{t}-\phi_{r}+\theta_{b}\right)/2$. It is noteworthy that, in well-planned localities, buildings often have similar orientation angles.
For such scenarios, we derive a closed form expression for the average number of first order reflection components. In this context, we note that the expression for $P_{b}/\phi_{b},\theta_{n}$ can be modified as $P_{b}/\phi_{b},\theta_{n}\approx (P_{self})^{i} \exp(-\lambda_{b}(d\left|\text{cos}\phi_{b}\right|(
\E[l]\text{tan}\theta_{n}+\E[w])+\E\left[l\right]\E\left[w\right])-\lambda_{h}W_{h}d|\text{cos}\phi_{b}|\text{sec}\theta_{n})$ with $i\in\{0,1,2\}$ by considering $\phi_{b}^{'}=0$ in (15) of \cite{rakesh2017analytical} and excluding Area($ZMC$) in the expression for $A_{block}$. To simplify the expression for $N_{r}$, the blocking probability $P_{b/\phi_{b},\theta_{n}}$ is approximated by a linear function with respect to tan$\theta_{n}$. We note that for most scenarios, tan$\theta_{n}$ is limited over a small range bounded by $\theta_{i}$ and $\theta_{u}$ (except at the values of $\theta_{n}$ which are close to 90$^0$).
Under such circumstances, the linearization of $P_{b/\phi_{b},\theta_{n}}$ will hold good. We show in the result section that this approximation results in negligible error when pointing direction of transmitter and receiver antenna beams are small.  The expression for $N_{r}$ conditioned on $\phi_{b}$ is obtained as,\\	\setcounter{equation}{8}
\begin{align}\label{N_r1}
N_{r/\phi_{b}}=&\frac{\lambda_{b}d(P_{self})^{i}}{2\left|\text{cos}\phi_{b}\right|^{-1}}\E_{L,W}\left[\int_{\theta_{i}}^{\theta_{u}}\frac{ae^{\left(-x \text{tan}\theta_{n}-y-z\text{sec}\theta_{n}\right)}\text{d}\theta_{n}}{\text{sec}^{-2}\theta_{n}}\right]\nonumber\\
\overset{\text{(a)}}{=}&\frac{\lambda_{b}d(P_{self})^{i}}{2\left|\text{cos}\phi_{b}\right|^{-1}}\E_{L,W}\hspace{-0.1cm}\left[\int_{\text{tan}\theta_{i}}^{\text{tan}\theta_{u}}\hspace{-0.45cm}ae^{\left(-x u-y-z\sqrt{1+u^{2}}\right)}\text{d}u\right]
\end{align}
where $u=\text{tan}\theta_{n}$ with $y=\lambda_{b}\left(d\left|\text{cos}\phi_{b}\right|\E[w]+\E\left[l\right]\E\left[w\right]\right)$, $x=\lambda_{b}d\left|\text{cos}\phi_{b}\right|\E[l]$, and $z=\lambda_{h}W_{h}d|\text{cos}\phi_{b}|$. We then apply a linear approximation to the function $e^{\left(-x u-y-z\sqrt{1+u^{2}}\right)}$ about the point $u=u_{0}=\frac{\text{tan}\theta_{i}+\text{tan}\theta_{u}}{2}$. Therefore, $N_{r}/\phi_{b}$ is approximated as,\\
\begin{align}\label{N_r2}
N_{r/\phi_{b}}\approx&\frac{\lambda_{b}d(P_{self})^{i}}{2\left|\text{cos}\phi_{b}\right|^{-1}}\E_{L,W}\left[\int_{\text{tan}\theta_{i}}^{\text{tan}\theta_{u}}\hspace{-0.2cm}
a\left[1+\left(x+\frac{2u_{0}z}{\sqrt{1+u_{0}^{2}}}\right) \right.\right.\nonumber\\&\left. \left.\hspace{0.9cm}\times (u_{0}-u)\vphantom{\int_{\text{sec}\theta_{i}}^{\text{sec}\theta_{u}}}\right]e^{\left(-xu_{0}-y-z\sqrt{1+u_{0}^{2}}\right)}\text{d}u\vphantom{\int_{\text{sec}\theta_{i}}^{\text{sec}\theta_{u}}}\right].
\end{align}
\begin{figure*}[!b]
	\hrule
	\vspace{0.2cm}
	\setcounter{equation}{13}
	\begin{align}\label{avg_pow}
	\frac{1}{PL}=\frac{1}{\pi}\left[\left(\int_{\phi_{i1}}
	^{\phi_{u1}}\hspace{-0.1cm}\frac{\text{d}
		\phi_{b}}{PL_{avg,1/\phi_{b}}}+\int_{\phi_{i2}}
	^{\phi_{u2}}\hspace{-0.1cm}\frac{\text{d}
		\phi_{b}}{PL_{avg,2/\phi_{b}}}\right)_{a=l}\hspace{-0.4cm}+\left(\int_{\phi_{i1}}
	^{\phi_{u1}}\hspace{-0.1cm}\frac{\text{d}
		\phi_{b}}{PL_{avg,1/\phi_{b}}}+\int_{\phi_{i2}}
	^{\phi_{u2}}\hspace{-0.1cm}\frac{\text{d}
		\phi_{b}}{PL_{avg,2/\phi_{b}}}\right)_{a=w}\right].
	\end{align} 
	\setcounter{equation}{15}
	\begin{align}\label{avg_sig_pow3}
	\frac{1}{PL_{\phi_{b}}}=&\frac{\Gamma_{l}\Gamma_{r,m}(P_{self})^{i}}{2d\lambda_{b}^{-1}\left|\text{cos}\phi_{b}\right|}\E_{L,W}\Bigg[a\bigg(\bigg(1+\bigg(x+\frac{2u_{0}z}{\sqrt{1+u_{0}^{2}}}\bigg)u_{0}\bigg)\left(\text{cos}\theta_{i}-\text{cos}\theta_{u}\right)+\bigg(x+\frac{2u_{0}z}{\sqrt{1+u_{0}^{2}}}\bigg)\bigg[\text{log}\left(\frac{\text{tan}\theta_{i}+\text{sec}\theta_{i}}{\text{tan}\theta_{u}+\text{sec}\theta_{u}}\right)\nonumber\\&\hspace{10cm}+\text{sin}\theta_{u}-\text{sin}\theta_{i}\bigg] \bigg)e^{-xu_{0}-y-z\sqrt{1+u_{0}^{2}}}\Bigg].
	\end{align}
\end{figure*} 
\setcounter{equation}{11}
The final expression for $N_{r/\phi_{b}}$ is derived in (\ref{N_r3}). 
\subsection{Evaluation of path loss in directional NLOS transmission}
In this section, we evaluate path loss for a typical directional NLOS link using the knowledge about the exact propagation trajectory based on the concept of image theory described earlier in section III-A, and thereafter calculating the length of each path.
We also take into account the loss due to reflection. Let us assume that $N^{i}$ number of buildings with orientation angle $\phi_{b}$ are present in the feasible region (discussed in Section III-A) in the $i$-th realization of the network. Hence, the overall directional path loss due to all reflections from buildings with orientation angle $\phi_{b}$ can be calculated as,\\
\begin{align}\label{camb}
	PL_{\phi_{b}}=\frac{G^{2}_{m}P_{t}}{P_{r,avg/\phi_{b}}}
=\frac{G^{2}_{m}P_{t}}{\sum_{N^{i}=1}^{\infty}{\sum_{n=1}^{N^{i}}}P_{r,n}P\left(N^{i}\right)}, 
\end{align}
where $P_{r,avg/\phi_{b}}$ denotes the average signal power received through the reflections, $P_{r,n}$ represents the power due to the $n$-th reflector, $P_{t}$ represents the transmit power, and $P\left(N^{i}\right)$ is the probability of occurrence of $N_{i}$ number of reflectors in the feasible region. We note that the computation of average received signal power is analogous to the local averaging of received signal power or PDP used to compute path loss in experimental evaluations \cite{rappaport201238}. Further, based on the discussion in Section III-D, the number of first order reflections is a Poisson random variable. The received signal power can be evaluated from Campbell's theorem, and therefore, (\ref{camb}) can be simplified as,\\
\begin{align}\label{avg_sig_pow1}
\frac{1}{PL_{\phi_{b}}}&=\frac{\Gamma_{l}\Gamma_{r}G_{m}^{2}P_{t}}{G^{2}_{m}P_{t}}\lambda_{b}
	\int_{A}\frac{\left(1-P_{b/\phi_{b},\theta_{n}}\right)}{\left(d\left|\text{cos}\phi_{b}\right|\text{sec}\theta_{n}\right)^{2}}\text{d}A\nonumber\\&=\Gamma_{l}\Gamma_{r}\lambda_{b}
\int_{A}\frac{\left(1-P_{b/\phi_{b},\theta_{n}}\right)}{\left(d\left|\text{cos}\phi_{b}\right|\text{sec}\theta_{n}\right)^{2}}\text{d}A,
\end{align}
where $\Gamma_{l}=\left(\frac{c}{4\pi f}\right)^{2}$ denotes the Friis free space path loss factor with $c$ and $f$ representing velocity of light and operating frequency, respectively. $\Gamma_{r}$ denotes the reflection loss. Experimental results \cite{rodriguez2015analysis} show that $\Gamma_{r}$ exhibits a `cosine' variation with respect to the angle of incidence $\theta_{in}$, i.e., $\Gamma_{r}=\Gamma_{r,m}\text{cos}\theta_{in}$, where $\Gamma_{r,m}$ denotes the maximum value of $\Gamma_{r}$ which is obtained for the normal incidence of the ray with the building surface.
In the proposed framework, $\Gamma_{r}=\Gamma_{r,m}\text{sin}\theta_{n}$ (see Fig.~\ref{fig:System Model}). Referring to Fig.~\ref{fig:Analytic1}(a), the total length of the reflected path corresponding to AoA $\theta_{n}$  is  $d\left|\text{cos}\phi_{b}\right|\text{sec}\theta_{n}$. The term $\left(1-P_{b/\phi_{b},\theta_{n}}\right)$ is the non-blocking probability for a reflected path with AoA, $\theta_{n}$. Moreover, the elemental area $\text{d}A$ can be found from (\ref{el_area}) given in Appendix A.
After substituting the value of $\text{d}A$ and the approximated expression for $P_{b/\phi,\theta_{n}}$ in (\ref{avg_sig_pow1}), we obtain (\ref{avg_pow}). For the scenarios with similar building orientation, we obtain closed form expression for path loss by the same approximation adopted in Section III-D. Therefore, we have,\\
\setcounter{equation}{14}
\begin{align}\label{avg_sig_pow2}
\frac{1}{PL_{\phi_{b}}}=&\frac{\Gamma_{l}\Gamma_{r,m}(P_{self})^{i}}{2d\lambda_{b}^{-1}\left|\text{cos}\phi_{b}\right|}\E_{L,W}\Bigg[\hspace{-0.15cm}\int_{\theta_{i}}^{\theta_{u}}\hspace{-0.15cm}\frac{ae^{\left(-x \text{tan}\theta_{n}-y-z\text{sec}\theta_{n}\right)}}{(\text{sin}\theta_{n})^{-1}}
\text{d}\theta_{n}\Bigg]\nonumber\\=&\frac{\Gamma_{l}\Gamma_{r,m}(P_{self})^{i}}{2d\lambda_{b}^{-1}\left|\text{cos}\phi_{b}\right|}\E_{L,W}\Bigg[\int_{\text{tan}\theta_{i}}^{\text{tan}\theta_{u}}\hspace{-0.1cm}\frac{ae^{\left(-xu_{0}-y-z\sqrt{1+u_{0}^{2}}\right)}}{(1+u^{2})^{\frac{3}{2}}}\nonumber\\&\times \left(1+\left(x+\frac{2u_{0}z}{\sqrt{1+u_{0}^{2}}}\right) (u_{0}-u)\right)\text{d}u\Bigg].
\end{align}
\setcounter{equation}{16}
The final expression for $PL_{\phi_{b}}$ is (\ref{avg_sig_pow3}).
Similar to the special case considered for the derivation of average number of first order reflections, the expression in (\ref{avg_sig_pow3}) can be used to determine the path loss experienced by the signal when the orientation angle for the buildings is the same. For buildings with different orientation angles, the path loss derived in (\ref{avg_pow}) can be used.
$PL_{\phi_{b},1}$ and $PL_{\phi_{b},2}$ denote the path loss corresponding to Scenario 1 and Scenario 2, respectively. The integration of the reciprocal of $PL_{\phi_{b},i}$, where $i\in\{1,2\}$ is equivalent to local averaging of signal power due to reflections from buildings with various orientation angles.
\section{Extension of the proposed framework to model second order reflections}
\begin{figure}[H]
	\centering
	\includegraphics[trim=1.7cm 0.2cm 1.6cm 0.2cm, clip=true, totalheight=0.30\textheight]{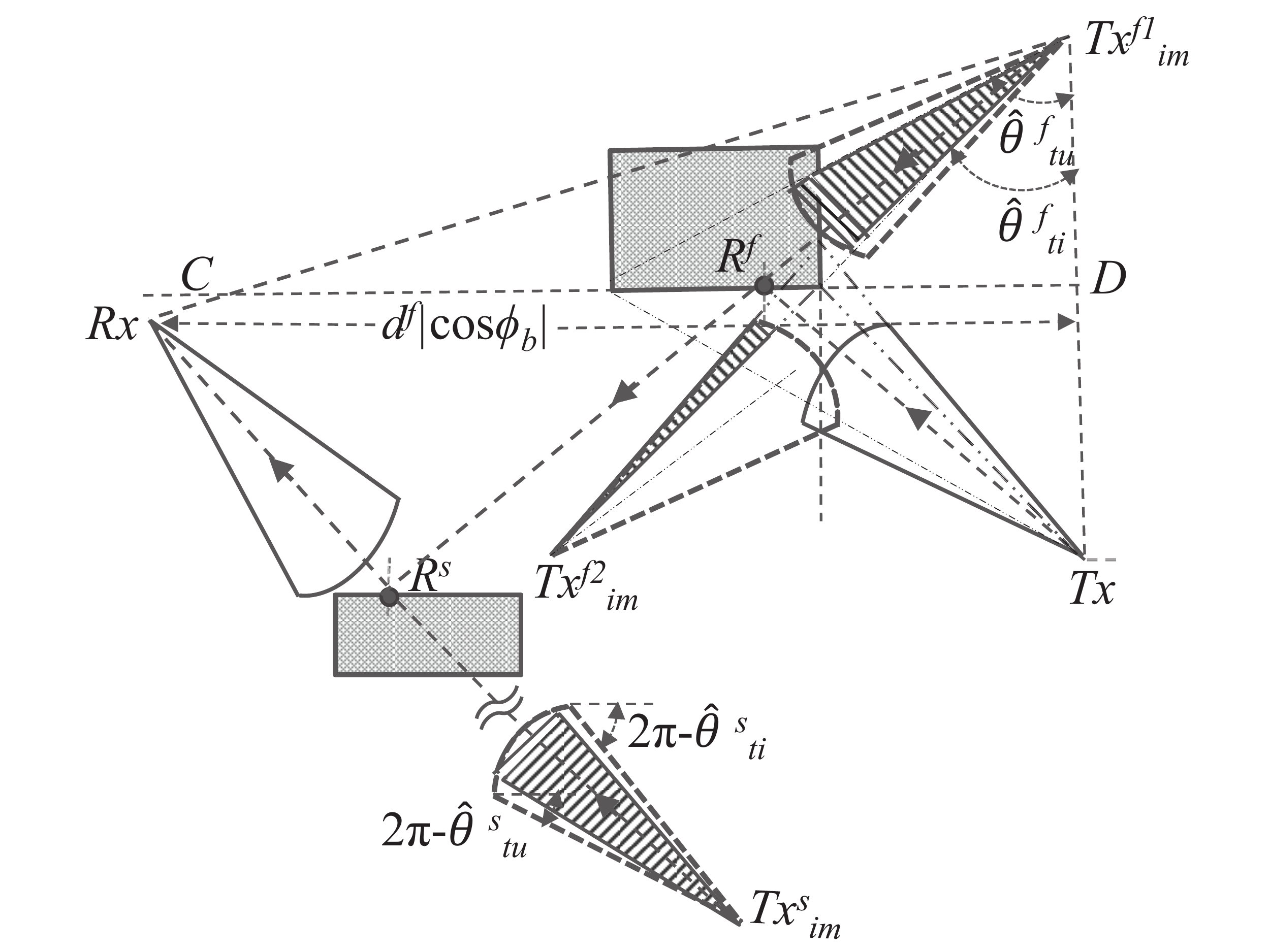}
	\caption{A deployment scenario which depicts second order reflection.}
	\label{fig:second_order_reflection}	
\end{figure}

Comparative studies of measurement campaigns and ray tracing simulations \cite{hur2016proposal,nguyen2014evaluation} reveal that a multi-path channel model consisting of first and few higher order (mostly second and third order) reflection components is sufficient to characterize most of the outdoor mmWave NLOS channels. In this section, we derive the mathematical expression for PDP of directional mmWave NLOS channels by combining statistics of first and second order reflection processes.
To achieve this goal, the proposed framework for modeling first order reflections is extended to second order reflections by decomposing each second order reflection into two consecutive first order reflections (corresponding point of reflections are denoted as $R^{f}$ and $R^{s}$) as depicted in Fig.~\ref{fig:second_order_reflection}. As a case study, we further assume the same orientation angle for all buildings; however, with additional computational complexity, the analysis can be extended to the general scenario of different building orientation angles.
As observed from Fig.~\ref{fig:second_order_reflection}, the second order reflection can be equivalently represented in terms of a first order reflection considering the image of the transmitter ($Tx_{im}^{f1}$) involved in the first reflection as a virtual transmitter for the second reflection. Thus, the main aim is to characterize the location and virtual antenna beam pattern (represented as the shaded portion within the antenna beam pattern in Fig.~\ref{fig:second_order_reflection}) of $Tx_{im}^{f1}$. We note that the statistical properties of $Tx_{im}^{f1}$ can readily be determined using the notion of a feasible region described in the following section.
\subsection{Characteristics of image transmitter resulting due to first reflection}
\begin{figure}[H]
	\centering
	\includegraphics[trim=2.8cm 2.4cm 1.6cm 5.5cm, clip=true, totalheight=0.21\textheight]{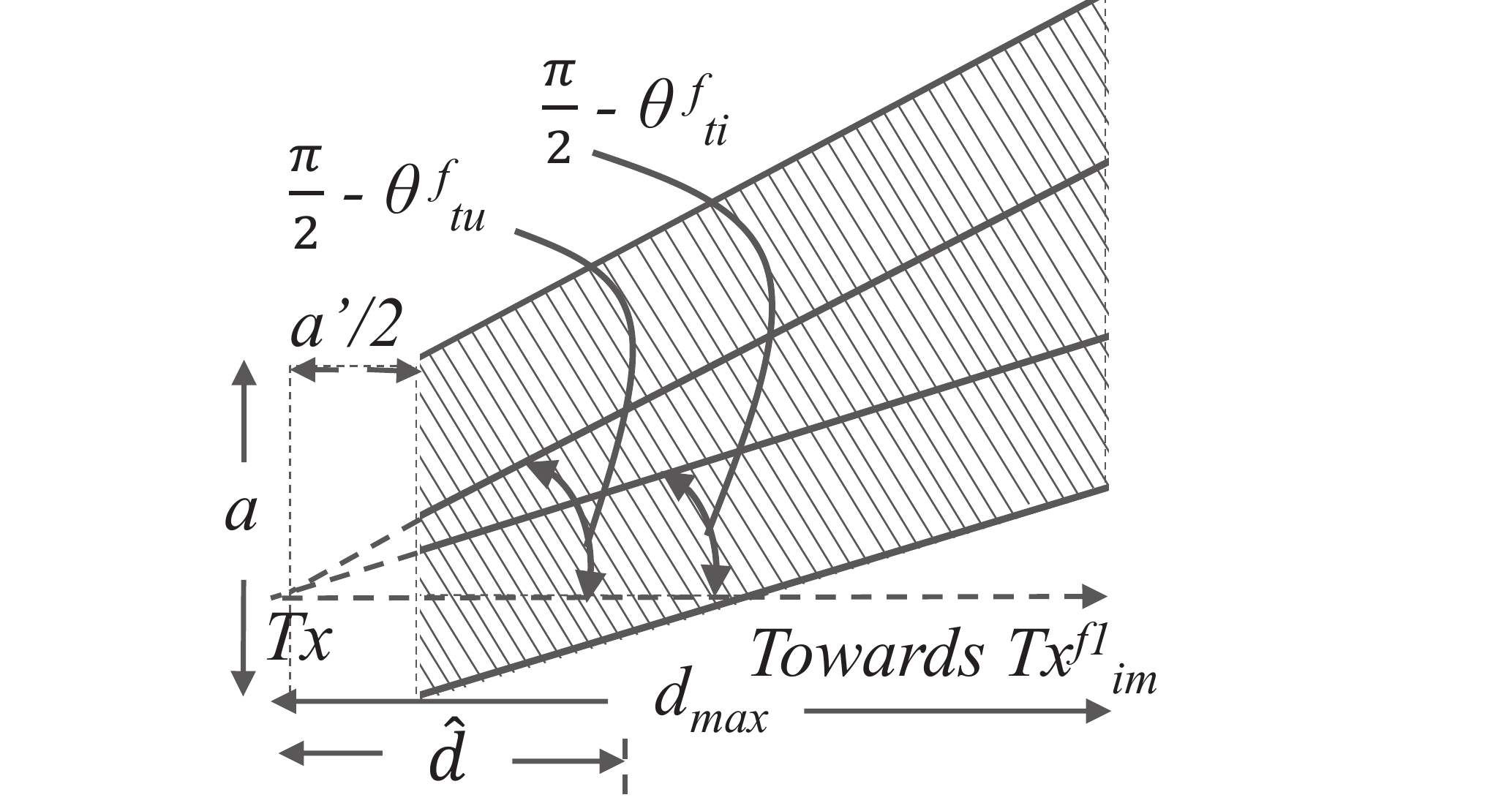}
	\caption{Feasible area for reflector location corresponding to first reflection.}
	\label{fig:feasible_area_second_order_reflection}	
\end{figure}

As reflectors are approximated by rectangular objects with random size and center point, the location and antenna radiation pattern of the image transmitter due to the first reflection is represented in terms of PDFs. For the sake of convenience, the distance and angular parameters introduced in Section II are redefined for first and second reflections by introducing the superscripts `$f$' and `$s$', respectively. For example, the first reflection is characterized by $d^{f}=d$, $\phi_{t}^{f}=\phi_{t}$, $\phi_{r}^{f}=\phi_{r}$, and $\theta_{n}^{f}=\theta_{n}$.
Depending on its location, each reflector creates at most two image transmitters (denoted by $Tx^{f1}_{im}$ and $Tx^{f2}_{im}$ as shown in Fig.~\ref{fig:second_order_reflection}).
Based on the location of the receiver node and pointing direction of its antenna beam, one of the image transmitters becomes the source for second reflection (the deployment scenario depicted in Fig.~\ref{fig:second_order_reflection} reveals that $Tx^{f1}_{im}$ is the candidate for second reflection). The statistical modeling of the image transmitter $Tx^{f1}_{im}$ with its virtual antenna beam pattern is considered in this section, and the same approach can be extended for $Tx^{f2}_{im}$ as well. Firstly, we identify the feasible region which will result in the first reflection from smooth surface of a reflector whose center lies within this region.
Conditioned on a maximum distance of $d_{max}=0.5d^{f}(|\text{cos}\phi_{b}|\text{tan}\theta_{ti}^{f}+|\text{sin}\phi_{b}|+\frac{a'}{2})$ for the location of the center of reflector from the point $Tx$ in the direction of $Tx_{im}^{f1}$ so that the reflected power is able to reach at the receiver, the feasible region is identified as the shaded area shown in Fig.~\ref{fig:feasible_area_second_order_reflection}. The variable $a'=w$ for $0\leq\phi_{b}\leq\frac{\pi}{2}$ and $a'=l$ for $\frac{\pi}{2}\leq\phi_{b}\leq\pi$. The area of the feasible region $A'$ is then evaluated as,
\begin{align}
A'=&\frac{a(d_{max}-\frac{a'}{2})}{2}\left(\frac{\text{sin}\theta_{ti}^{f}}{\text{cot}\theta_{ti}^{f}}+\frac{\text{sin}\theta_{tu}^{f}}{\text{cot}\theta_{tu}^{f}}\right)+\frac{d_{max}^{2}-\left(\frac{a'}{2}\right)^{2} }{2}\nonumber\\&\hspace{4cm}\times\left(\text{cot}\theta_{tu}^{f}-\text{cot}\theta_{ti}^{f}\right).
\end{align}
The dependency of PDP on reflector location can be obtained from Cambpell's theorem using a similar approach adopted in Section III-E. However, this approach is highly complex for a second order reflection and therefore, we proceed the derivation of PDP by making an assumption that at most one reflector can exist in the feasible region $A'$ for a given realization of the scenario.
This is true for most building densities and dimensions since probability of existence of multiple buildings in $A'$ is negligible. Hence, the occurrence of first reflection can be modeled using a Bernoulli random variable $\Pi$ with success probability $p=1-\exp(-\lambda_{b}A')$.
To model the virtual antenna beam pattern of $Tx^{f1}_{im}$, we derive the angular parameters ($\hat{\theta}_{ti}^{f}$ and $\hat{\theta}_{tu}^{f}$) which define the edges of the virtual antenna beam pattern (see in Fig.~\ref{fig:second_order_reflection}). Due to the dependency between $\hat{\theta}_{ti}^{f}$ and $\hat{\theta}_{tu}^{f}$, the statistics of $\hat{\theta}_{ti}^{f}$ is governed by a joint PDF, $f_{\hat{D},\hat{\Theta}_{ti}^{f}}\left(\hat{d},\hat{\theta}_{ti}^{f}\right)$, whereas $\hat{\theta}_{tu}^{f}$ is modeled as a deterministic  function of $\hat{\theta}_{ti}^{f}$.
The joint PDF, $f_{\hat{D},\hat{\Theta}_{ti}^{f}}\left(\hat{d},\hat{\theta}_{ti}^{f}\right)$ can be written as, $f_{\hat{D},\hat{\Theta}_{ti}^{f}}\left(\hat{d},\hat{\theta}_{ti}^{f}\right)=f_{\hat{\Theta}_{ti}^{f}}\left(\hat{\theta}_{ti}^{f}/\hat{d}\right)f_{\hat{D}}(\hat{d})$, where $f_{\hat{D}}(\hat{d})$ and $f_{\hat{\Theta}_{ti}^{f}}\left(\hat{\theta}_{ti}^{f}/\hat{d}\right)$ is evaluated (detailed derivation is provided in Appendix B) using Fig.~\ref{fig:second_order_reflection} and Fig.~\ref{fig:feasible_area_second_order_reflection} as,\\
\begin{align}
f_{\hat{D}}(\hat{d})=&\frac{C_{1}+2C_{2}\hat{d}}{C_{1}\left( d_{max}-\frac{a'}{2}\right)+C_{2}\left(d_{max}^{2}-\left(\frac{a'}{2}\right)^{2}\right)},\nonumber\\ &\hspace{4cm}\frac{a'}{2}\leq\hat{d}\leq d_{max}.
\end{align}
\begin{align}\label{PDF_theta_n_final}
f_{\hat{\Theta}_{ti}^{f}}(\hat{\theta}_{ti}^{f}/\hat{d})=\left\{\begin{array}{lr}
\hspace{-0.25cm}\frac{\text{sec}^{2}\hat{\theta}_{ti}^{f}}{\text{tan}(\frac{\pi}{2}-\theta_{tu}^{f})-\text{tan}\theta_{li}^{f}}, &\hspace{-0.2cm} \frac{\pi}{2}-\theta_{ti}^{f}\leq\hat{\theta}_{ti}^{f}\leq\frac{\pi}{2}-\theta_{tu}^{f},\\\\
\hspace{-0.25cm}\frac{\text{tan}(\frac{\pi}{2}-\theta_{ti}^{f})-\text{tan}\theta_{li}^{f}}{\text{tan}(\frac{\pi}{2}-\theta_{tu}^{f})-\text{tan}\theta_{li}^{f}}, & \hat{\theta}_{ti}^{f}=\frac{\pi}{2}-\theta_{ti}^{f},\\\\
0, & \text{otherwise.}
\end{array}\right.
\end{align}
The expression for $\hat{\theta}_{tu}^{f}$ can be derived using Fig.~\ref{fig:second_order_reflection} and Fig.~\ref{fig:feasible_area_second_order_reflection} as, $\hat{\theta}_{tu}^{f}=\text{tan}^{-1}\left(\hat{\theta}_{ti}^{f}+\frac{a}{\hat{d}-a'/2}\right)$, for $\frac{\pi}{2}-\theta_{ti}^{f}\leq\hat{\theta}_{tu}^{f}\leq\frac{\pi}{2}-\theta_{tu}^{f}$; $\theta_{tu}^{f}$, for $\hat{\theta}_{tu}^{f}>\theta_{tu}^{f}$.
Based on the preceding formulation, we can evaluate certain metrics related to  directional NLOS mmWave channels such as average number of reflection components, path loss, PDP, etc.
\begin{figure*}[!b]
	\hrule
	\vspace{0.2cm}
	\setcounter{equation}{27}
	\begin{align}\label{int_1}
	\int_{0}^{\infty}\hspace{-0.2cm}PDP(\tau)\text{d}\tau
	=&\Bigg[\frac{\Gamma_{l}\Gamma_{r,m}\lambda_{b}}{2d^{f}\left|\text{cos}\phi_{b}\right|}\E_{L,W}\Bigg[a\Bigg((1+\overline{x}^{f}u_{0}^{f})\left[\text{cos}\theta_{i}^{f}-\text{cos}\theta_{u}^{f}\right]+\overline{x}^{f}\left[\text{log}\left(\frac{\text{tan}\theta_{i}^{f}+\text{sec}\theta_{i}^{f}}{\text{tan}\theta_{u}^{f}+\text{sec}\theta_{u}^{f}}\right)  +\text{sin}\theta_{u}^{f}-\text{sin}\theta_{i}^{f}\right] \Bigg)\nonumber\\&\times e^{-x^{f}u_{0}^{f}-y^{f}-z^{f}\sqrt{1+(u_{0}^{f})^{2}}}\Bigg]+
	\frac{\Gamma_{l}\left(\Gamma_{r,m}\right)^{2}\lambda_{b}}{2d^{f}\left|\text{cos}\phi_{b}\right|}\int_{\hat{d}}\int_{\hat{\theta}_{ti}^{f}}\E_{L,W}\Bigg[a\bigg((1+\overline{x}^{s}u_{0}^{s})\left(\frac{\theta_{u}^{s}-\theta_{i}^{s}}{2} -\frac{\text{sin}2\theta_{u}^{s}-\text{sin}2\theta_{i}^{s}}{4}\right) \nonumber\\&-\overline{x}^{s}\left(\text{log}\left(\frac{\text{cos}\theta_{i}^{s}}{\text{cos}\theta_{u}^{s}}\right)  +\frac{\text{cos}^{2}\theta_{u}^{s}-\text{cos}^{2}\theta_{i}^{s}}{2}\right) \bigg)e^{-x^{f}u_{0}^{s}-y^{f}-z^{f}\sqrt{1+(u_{0}^{s})^{2}}}\Bigg]p f_{\hat{D},\hat{\Theta}_{ti}^{f}}\left(\hat{d},\hat{\theta}_{ti}^{f}\right)\text{d}\hat{\theta}_{ti}^{f}\text{d}\hat{d}\Bigg](P_{self})^{i}.
	\end{align}
	\begin{align}\label{int_2}
	\int_{0}^{\infty}\hspace{-0.2cm}\tau PDP(\tau)\text{d}\tau
	=&\Bigg[\frac{\Gamma_{l}\Gamma_{r,m}}{2c\lambda_{b}^{-1}}\E_{L,W}\Bigg[a\Bigg((1+\overline{x}^{f}u_{0}^{f})\text{log}\left(\frac{\text{cos}\theta_{i}^{f}}{\text{cos}\theta_{u}^{f}}\right)-\overline{x}^{f}\left(\text{tan}\theta_{u}^{f}-\text{tan}\theta_{i}^{f}+\theta_{i}^{f}-\theta_{u}^{f}\right) \Bigg)e^{-x^{f}u_{0}^{f}-y^{f}-z^{f}\sqrt{1+(u_{0}^{f})^{2}}}\Bigg]\nonumber\\
	&+\int_{\hat{d}}\int_{\hat{\theta}_{ti}^{f}}\frac{\Gamma_{l}(\Gamma_{r,m})^{2}}{2c\lambda_{b}^{-1}}\E_{L,W}\Bigg[a\Bigg((1+\overline{x}^{s}u_{0}^{s})\left[\text{log}\left(\frac{\text{tan}\theta_{u}^{s}+\text{sec}\theta_{u}^{s}}{\text{tan}\theta_{i}^{s}+\text{sec}\theta_{i}^{s}}\right)  +\text{sin}\theta_{u}^{s}-\text{sin}\theta_{i}^{s} \right]\nonumber\\&-\overline{x}^{s}\left[\frac{\text{sec}^{2}\theta_{u}^{s}+1}{\text{sec}\theta_{u}^{s}}-\frac{\text{sec}^{2}\theta_{i}^{s}+1}{\text{sec}\theta_{i}^{s}}\right] \Bigg)e^{-x^{f}u_{0}^{s}-y^{f}-z^{f}\sqrt{1+(u_{0}^{s})^{2}}}\Bigg]p f_{\hat{D},\hat{\Theta}_{ti}^{f}}\left(\hat{d},\hat{\theta}_{ti}^{f}\right)\text{d}\hat{\theta}_{ti}^{f}\text{d}\hat{d}\Bigg](P_{self})^{i}.
	\end{align}
	\begin{align}\label{int_3}
	\int_{0}^{\infty}\hspace{-0.2cm}\tau^{2} PDP(\tau)\text{d}\tau
	=&(P_{self})^{i}\Bigg[\frac{\Gamma_{l}\Gamma_{r,m}\lambda_{b}d^{f}}{2c^{2}\left|\text{cos}\phi_{b}\right|^{-1}}\E_{L,W}\Bigg[a\Bigg((1+\overline{x}^{f}u_{0}^{f})\left[\text{sec}\theta_{u}^{f}-\text{sec}\theta_{i}^{f}\right]-\frac{\overline{x}^{f}}{2}\Bigg[\text{tan}\theta_{u}^{f}\text{sec}\theta_{u}^{f}-\text{log}\left(\frac{\text{tan}\theta_{u}^{f}+\text{sec}\theta_{u}^{f}}{\text{tan}\theta_{i}^{f}+\text{sec}\theta_{i}^{f}}\right)\nonumber\\&-\text{tan}\theta_{i}^{f}\text{sec}\theta_{i}^{f}\Bigg] \Bigg)e^{-x^{f}u_{0}^{f}-y^{f}-z^{f}\sqrt{1+(u_{0}^{f})^{2}}}\Bigg]+\frac{\Gamma_{l}(\Gamma_{r,m})^{2}\lambda_{b}d^{f}}{2c^{2}\left|\text{cos}\phi_{b}\right|^{-1}}\int_{\hat{d}}\int_{\hat{\theta}_{ti}^{f}}\hspace{-0.25cm}\E_{L,W}\Bigg[a\Bigg((1+\overline{x}^{f}u_{0}^{s})\Big[\theta_{i}^{s}-\theta_{u}^{s}+\text{tan}\theta_{u}^{s}\nonumber\\&-\text{tan}\theta_{i}^{s}\Big]-\overline{x}^{f}\Bigg[\frac{\text{tan}^{2}\theta_{u}^{s}-\text{tan}^{2}\theta_{i}^{s}}{2}+\text{log}\left(\frac{\text{sec}\theta_{i}^{s}}{\text{sec}\theta_{u}^{s}}\right)\Bigg] \Bigg)e^{-x^{f}u_{0}^{s}-y^{f}-z^{f}\sqrt{1+(u_{0}^{s})^{2}}}\Bigg]p f_{\hat{D},\hat{\Theta}_{ti}^{f}}\left(\hat{d},\hat{\theta}_{ti}^{f}\right)\text{d}\hat{\theta}_{ti}^{f}\text{d}\hat{d}\Bigg].
	\end{align}
	\setcounter{equation}{19}
\end{figure*} 
\subsection{Power delay profile of directional NLOS channel}
The statistical independence of reflection components can be leveraged to represent the PDP of a directional NLOS mmWave channel as, $PDP(\tau)=PDP^{f}(\tau)+PDP^{s}(\tau)$, where $PDP^{j}(\tau)$ denotes the PDP due to $j$-th order ($i\in\{f,s\}$) reflection and $\tau$ denotes the delay variable.
Additionally, we define $\overline{PDP^{j}}(\tau)=P^{j}(\tau)f^{j}(\tau)$, where $P^{j}(\tau)$ and $f^{j}(\tau)$ represent the received power density with respect to delay $\tau$ and number of reflection components arriving at the receiver with delay $\tau$, respectively. $f^{j}(\tau)$ can be evaluated based on (\ref{f_t}) as,\\
\begin{align}\label{f_t}
f^{j}(\tau)=\begin{array}{lr}
\text{lim}\\
|\tau_{2}-\tau_{1}|\rightarrow 0\end{array} \frac{\E\left(N^{j}|\tau_{1}\leq \tau\leq\tau_{2}\right)}{|\tau_{2}-\tau_{1}|},
\end{align}
where $N^{j}$ denotes the number of $j$-th order reflection components arriving within the interval $\tau_{1}\leq\tau\leq\tau_{2}$. Delay parameters $\tau_{1}$ and $\tau_{2}$ correspond to propagation distances $d^{j}_{1}$ and $d^{j}_{2}$ ($\tau_{1}=d^{f}|\text{cos}\phi_{b}|\text{sec}\theta_{1}^{j}$ and $\tau_{2}=d^{f}|\text{cos}\phi_{b}|\text{sec}\theta_{2}^{j}$), respectively. Let $\theta^{j}_{1}=\theta^{j}_{n}$ and $\theta^{j}_{2}=\theta^{j}_{n}+\Delta\theta^{j}_{n}$. Hence, substitution of various parameters in (\ref{f_t}) results in,\\
\begin{align}\label{f_tau}
f^{j}(\tau)=\begin{array}{lr}
\text{lim}\\
|\Delta\theta_{n}^{j}|\rightarrow 0\end{array}\frac{\mathcal{N}}{\mathcal{D}},
\end{align}
where  $\mathcal{N}=0.5a\lambda_{b}d^{f}|\text{cos}\phi_{b}|\left[\text{tan}(\theta_{n}^{j}+\Delta\theta_{n}^{j})-\text{tan}\theta_{n}^{j}\right]P_{b/\phi_{b},\theta_{n}^{j}}$, $\mathcal{D}=c^{-1}d^{f}|\text{cos}\phi_{b}|\left[\text{sec}(\theta_{n}^{j}+\Delta\theta_{n}^{j})-\text{sec}\theta_{n}^{j}\right]$.  $P_{b/\phi_{b},\theta_{n}^{j}}\approx(P_{self})^{i}\exp(-\lambda_{b}(d^{j}\left|\text{cos}\phi_{b}\right|(
\E[l]\text{tan}\theta^{j}_{n}+\E[w])+\E\left[l\right]\E\left[w\right])-\lambda_{h}W_{h}d^{j}|\text{cos}\phi_{b}|\text{sec}\theta^{j}_{n})$ with $i\in\{0,1,2\}$. Using trigonometric identities, (\ref{f_tau}) can be rewritten as,\\
\begin{align}
f^{j}(\tau)=\begin{array}{lr}
\text{lim}\\
|\Delta\theta_{n}^{j}|\rightarrow 0\end{array}\frac{ac\lambda_{b}\text{sin}\Delta\theta_{n}^{j}P_{b/\phi_{b},\theta_{n}^{j}}}{2\left(\text{cos}\theta_{n}^{j}-\text{cos}(\theta_{n}^{j}+\Delta\theta_{n}^{j})\right)}
\end{align}
and applying L'-Hospital's rule, we obtain,\\
\begin{align}
f^{j}(\tau)=\frac{ac\lambda_{b}P_{b/\phi_{b},\theta_{n}^{j}}}{2\text{sin}\theta_{n}^{j}}
\end{align}
Based on the Friis free space equation for path loss, $P^{j}(\tau)=\Gamma_{l}\Gamma_{r}^{j}(d^{f}|\text{cos}\phi_{b}|\text{sec}\theta_{n}^{j})^{-2}=\Gamma_{l}\Gamma_{r}^{j}(c\tau)^{-2}$. Therefore, the expression for $\overline{PDP^{j}}(\tau)$ is obtained as,\\
\begin{align}
\overline{PDP^{j}}(\tau)=&\frac{a\Gamma_{r}^{j}\Gamma_{l}(P_{self})^{i}}{2\lambda_{b}^{-1}c\tau^{2}\text{sin}\theta_{n}^{j}} \exp(-\lambda_{b}(d^{j}\left|\text{cos}\phi_{b}\right|(
\E[l]\text{tan}\theta^{j}_{n}+\nonumber\\&\E[w])+\E\left[l\right]\E\left[w\right])-\lambda_{h}W_{h}d^{j}|\text{cos}\phi_{b}|\text{sec}\theta^{j}_{n})
\end{align}
where $\text{tan}\theta_{n}^{j}=\sqrt{\left(\frac{c\tau}{d^{f}|\text{cos}\phi_{b}|}\right)^{2}-1}$,  $\Gamma_{r}^{f}=\Gamma_{r,m}\text{sin}\theta^{f}_{n}$, $\text{sin}\theta_{n}^{s}=\sqrt{1-\left(\frac{d^{f}|\text{cos}\phi_{b}|}{c\tau}\right)^{2}}$,  $\Gamma_{r}^{s}=(\Gamma_{r,m})^{2}\text{sin}^{2}\theta^{s}_{n}$ (owing to the fact that $\theta_{n}^{f}=\theta_{n}^{s}$), $\text{sec}\theta^{j}_{n}=\frac{c\tau}{d^{f}|\text{cos}\phi_{b}|}$. The PDP of directional channel is given by,\\
\begin{align}
PDP^{j}(\tau)=\left\{\hspace{-0.2cm}\begin{array}{lr}
\overline{PDP^{f}}(\tau)I(\tau^{f}_{i},\tau^{f}_{u}), & j=f\\\\
\int_{\hat{d}}\int_{\hat{\theta}_{tl}^{f}}\overline{PDP^{s}}(\tau)I(\tau^{s}_{i},\tau^{s}_{u})\\\hspace{0.5cm}\times p f_{\hat{D},\hat{\Theta}_{tl}^{f}}\left(\hat{d},\hat{\theta}_{tl}^{f}\right)\text{d}\hat{\theta}_{tl}^{f}\text{d}\hat{d}, & j=s
\end{array}\right.
\end{align}
where $I(\tau^{j}_{i},\tau^{j}_{u})=1,$ for $\tau_{i}^{j}\leq\tau\leq\tau_{u}^{j};0,$ otherwise. The parameters $\tau_{i}^{j}$ and $\tau_{u}^{j}$ are evaluated as, $\tau_{i}^{j}=\frac{d^{f}|\text{cos}\phi_{b}|\text{sec}\theta_{i}^{j}}{c}$ and $\tau_{u}^{j}=\frac{d^{f}|\text{cos}\phi_{b}|\text{sec}\theta_{u}^{j}}{c}$ with $\theta_{i}^{j}$ and $\theta_{u}^{j}$ calculated based on the scenarios discussed in Section III-B. Finally, average delay ($\overline{\tau}$) and RMS delay spread ($\sigma_{\tau}$) are calculated from the following formulas,\\
\begin{align}\label{avg_del}
\overline{\tau}=\frac{\int_{0}^{\infty}\tau PDP(\tau)\text{d}\tau}{\int_{0}^{\infty}PDP(\tau)\text{d}\tau}
\end{align}
\begin{align}\label{rms_del}
\sigma_{\tau}=\sqrt{\frac{\int_{0}^{\infty}(\tau-\overline{\tau})^{2} PDP(\tau)\text{d}\tau}{\int_{0}^{\infty}PDP(\tau)\text{d}\tau}}=\sqrt{\overline{\tau^{2}}-\overline{\tau}^{2}},
\end{align}
where $\overline{\tau^{2}}=\left(\int_{0}^{\infty}\tau^{2} PDP(\tau)\text{d}\tau\right)/\left(\int_{0}^{\infty}PDP(\tau)\text{d}\tau\right)$. The integrals in (\ref{avg_del}) and (\ref{rms_del}) can be computed using the substitution $\tau=d^{f}|\text{cos}\phi_{b}|\text{sec}\theta_{n}^{j}$, where $j\in\{f,s\}$, and we apply the same linear approximation for blocking probability used for the evaluation of average number of first order reflections and path loss in Section III-D and Section III-E, respectively. Let, $x^{f}=\lambda_{b}d^{f}|\text{cos}\phi_{b}|\E[l]$, $y^{f}=\lambda_{b}(d^{f}|\text{cos}\phi_{b}|\E[w]+\E[l]\E[w])$, $u_{0}^{s}=(\text{tan}\theta_{i}^{s}+\text{tan}\theta_{u}^{s})/2$,
$\overline{x}^{f}=x^{f}+\frac{2u_{0}^{f}z^{f}}{\sqrt{1+(u_{0}^{f})^{2}}}$, and $\overline{x}^{s}=x^{f}+\frac{2u_{0}^{s}z^{f}}{\sqrt{1+(u_{0}^{s})^{2}}}$
We note that the angular parameters $\theta_{i}^{s}$ and $\theta_{u}^{s}$ are a function of $\hat{\theta_{ti}^{f}}$ and $\hat{\theta_{tu}^{f}}$ depending on the scenarios considered in Section III-B. Although, the derived expressions for $\overline{\tau}$ and $\sigma_{\tau}$ involve integrals, they do provide ready estimates of the delay parameters of a directional NLOS channel as compared to a ray tracing based simulation which is rather cumbersome and time consuming.
Moreover, it should be noted that a closed form expression for both $\overline{\tau}$ and $\sigma_{\tau}$ can be obtained if only first order reflection components are considered in (\ref{int_1}), (\ref{int_2}), and (\ref{int_3}) (evaluation with only $j=f$).
\begin{figure*}
	\centering
	\includegraphics[height=4.7cm,width=4.8cm]{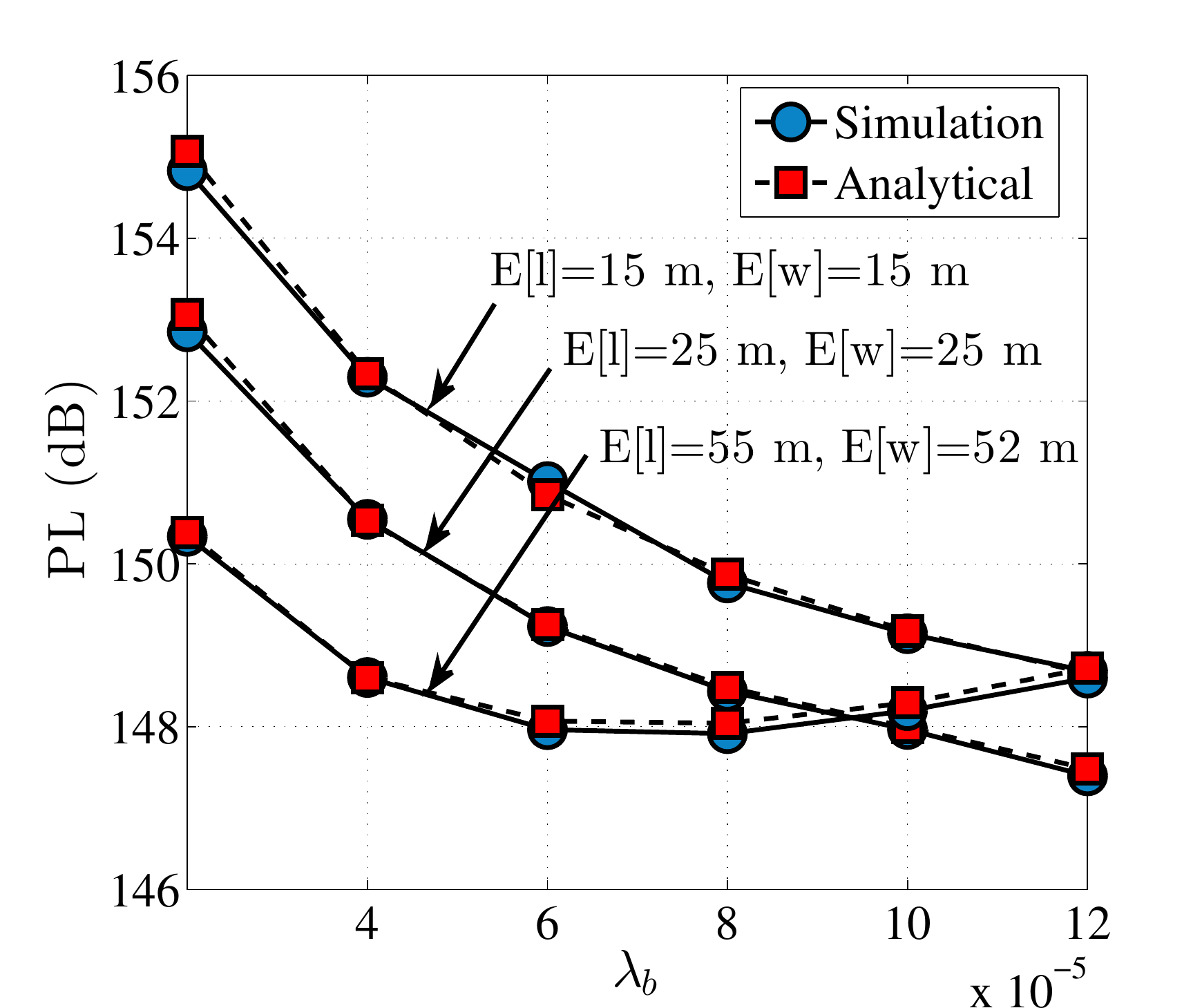}\hspace{-1.3em}
	\includegraphics[height=4.7cm,width=4.8cm]{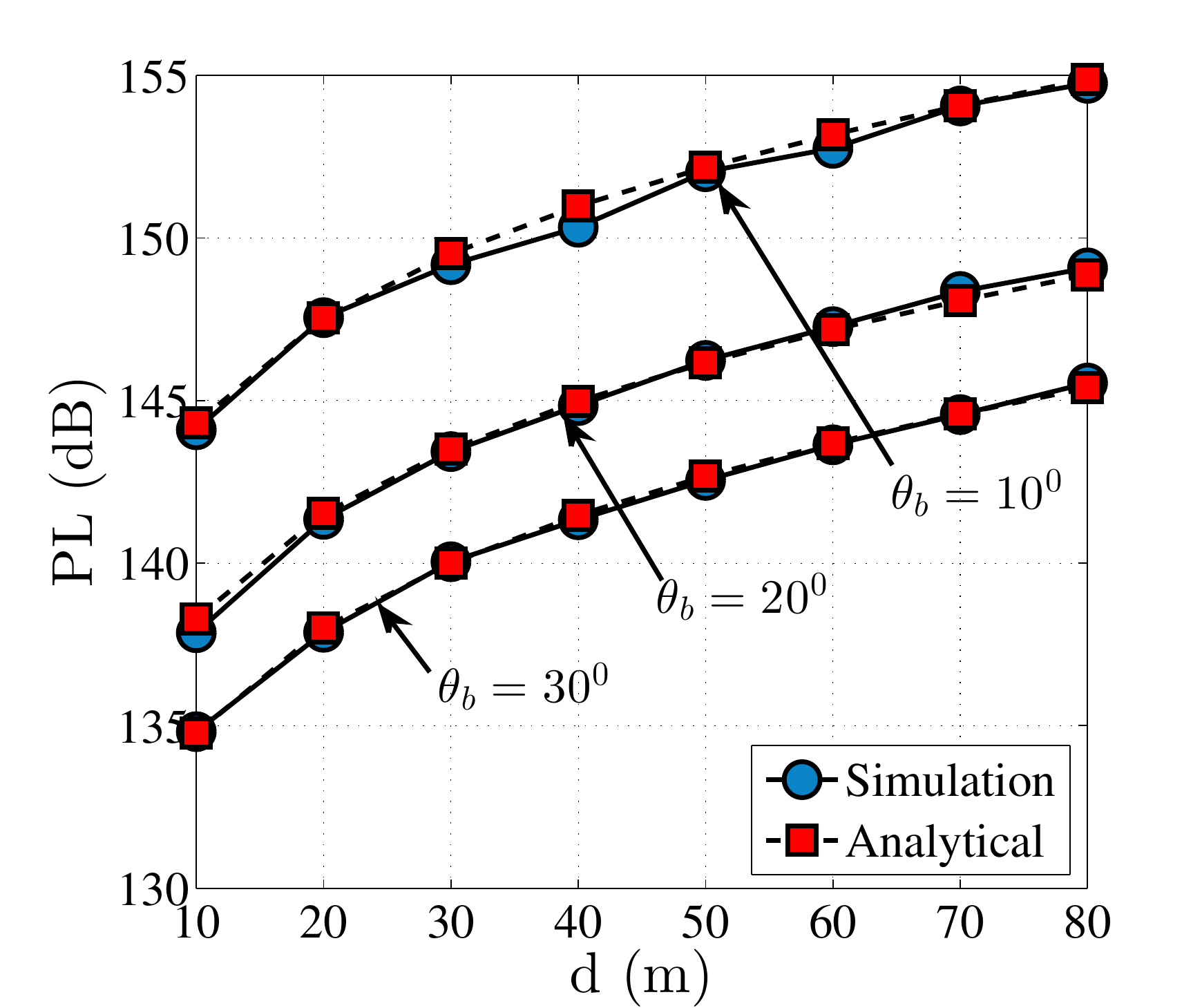}\hspace{-0.75em}
	\includegraphics[height=4.7cm,width=4.7cm]{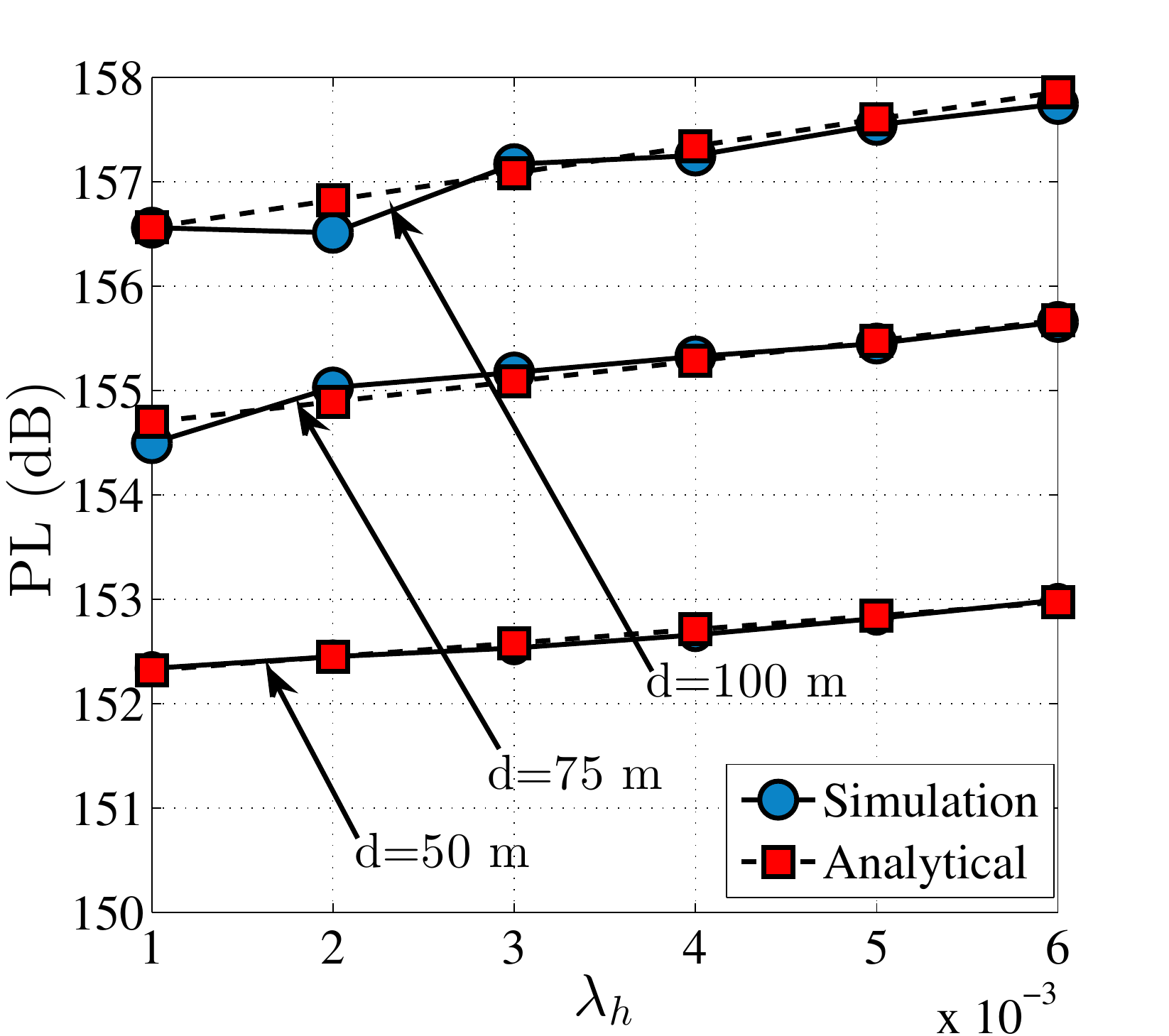}
	{$\mspace{1mu}$(a)$\mspace{220mu}$(b)$\mspace{220mu}$(c)}          
	\caption{Path loss for variable (a) $\lambda_{h}$, E[$l$], and E[$w$] with $\theta_{b}=20^{0}$ and $d=75$ m, (b) $d$ and $\theta_{b}$ with $\lambda_{b}=8\times 10^{-5}$ and $\E[l]=\E[w]=25$ m, (c) $\lambda_{h}$ and $d$ with $\lambda_{b}=8\times 10^{-5}$, $\theta_{b}=20^{0}$, $W_{h}=30$ cm, $P_{self}=0.25$, and $\E[l]=\E[w]=25$ m.}
	\label{fig:performance_analysis}
	\vspace{-0.25cm}
\end{figure*}
\section{Performance Analysis and Discussion}
In this section, we investigate the accuracy of the proposed geometry based channel model by comparing the analytical solution with the results obtained through simulation. We also draw a comparison between data predictions based on our model with that reported by credible measurement campaigns. In addition, we present an insight into the performance of directional NLOS mmWave links affected by blockages. For simulation purposes, we assume that the receiver is located at the center in a 800$\times$800 m$^{2}$ area, and separated from the transmitter by a distance $d$. Operating frequency and transmit power are chosen as
$38$ GHz and $0$ dBW, respectively. Antenna beam pointing angle for the transmitter and receiver are arbitrarily chosen as $\phi_{t}=110^{0}$ and $\phi_{r}=50^{0}$. The valid reflections are found after ascertaining that the reflection coming from the edge of the reflector is captured within the receiver main lobe. The maximum reflection loss is taken as 19.1 dB \cite{rodriguez2015analysis}. Building density $\lambda_{b}$ varies from $2\times10^{-5}$ to $12\times10^{-5}$ which is typical in outdoor environments \cite{rappaport201238,bai2014analysis}. Moreover, the average size of buildings is chosen from $15\times15$ m$^{2}$ to $55\times55$ m$^{2}$ \cite{bai2014analysis} to enable modeling a wide variety of buildings. It should be noted that the results presented in this section are based on the assumption that $\theta_{b,t}=\theta_{b,r}=\theta_{b}$, and averaged over 2$\times$10$^{5}$ realizations of the PPP.
\subsection{Validation of the proposed channel model based on comparison of simulation and experimental data}
In this section, we analytically evaluate channel parameters and compare them with the results generated by extensive simulations. We investigate the effect of building density on path loss by specifying the average building size, and considering only first order reflection components. Simulation results in Fig.~\ref{fig:performance_analysis}(a) show that the path loss decreases as the density of buildings in the deployment area increases. This can be attributed to the increase in number of reflection components. Moreover, it is also observed that larger buildings provide more number of reflection components than buildings with smaller dimensions.
Interestingly, for very high building density, the path loss begins to increase if the building dimensions are large. This happens since the probability of blockage of reflection components is more for large sized buildings - assuming the same building density, smaller buildings create more open spaces for the reflected components and therefore lead to fewer number of blockage events.
We also evaluate the effect of the transmission distance on the path loss (Fig.~\ref{fig:performance_analysis}(b)) with varying $\theta_{b}$. As depicted in Fig.~\ref{fig:performance_analysis}(b), the path loss increases monotonically with distance which is a known fact.  Moreover, the path loss reduces for larger antenna HPBW. This is due to the fact that the nodes with wider antenna beam pattern are able to receive a greater number of multiple reflection components simultaneously.
We proceed to investigate the effect of human blockage of refection components in Fig.~\ref{fig:performance_analysis}(c) by varying the parameters $\lambda_{h}$ and $d$. We consider a scenario where one of the nodes is carried by a person and the other node is placed in a fixed location. The plots in  Fig.~\ref{fig:performance_analysis}(c) reveal that a human density variation from $1\times10^{-3}$ to $6\times10^{-3}$ introduces a nominal change of  $1\sim 2$ dB in path loss. We also observe that the analytically evaluated values of path loss across various operating environments shows a good match with corresponding simulated values of path loss.
 \begin{figure}[H]
 	\centering
 	\begin{subfigure}{0.23\textwidth}
 		\centering
 		\includegraphics[height=4.7cm,width=4.7cm]{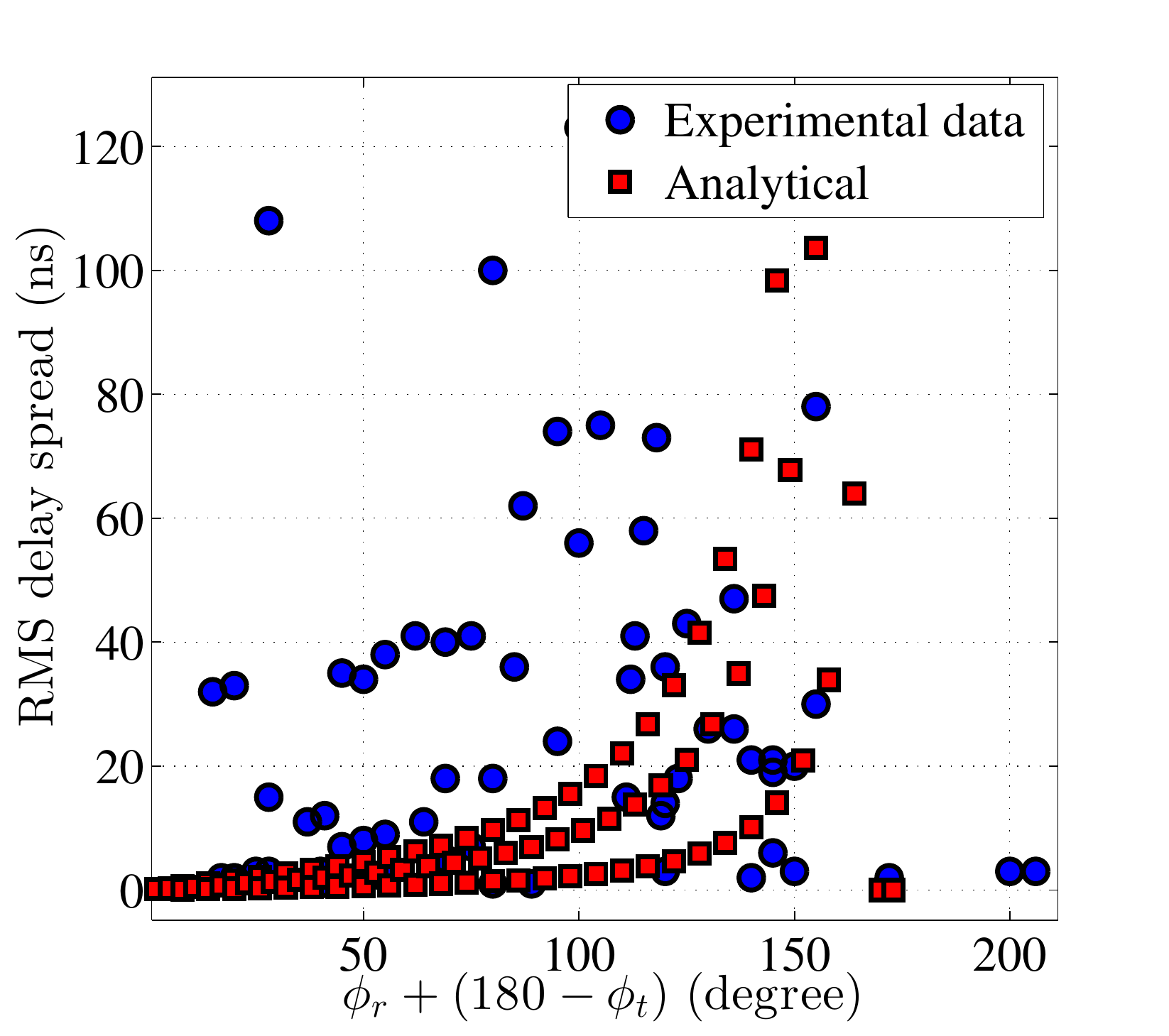}
 		\vspace{-0.65cm}
 		\caption{}
 	\end{subfigure}
 	\hspace{-0.025cm}
 	\begin{subfigure}{0.22\textwidth}
 		\centering
 		\includegraphics[height=4.7cm,width=4.7cm]{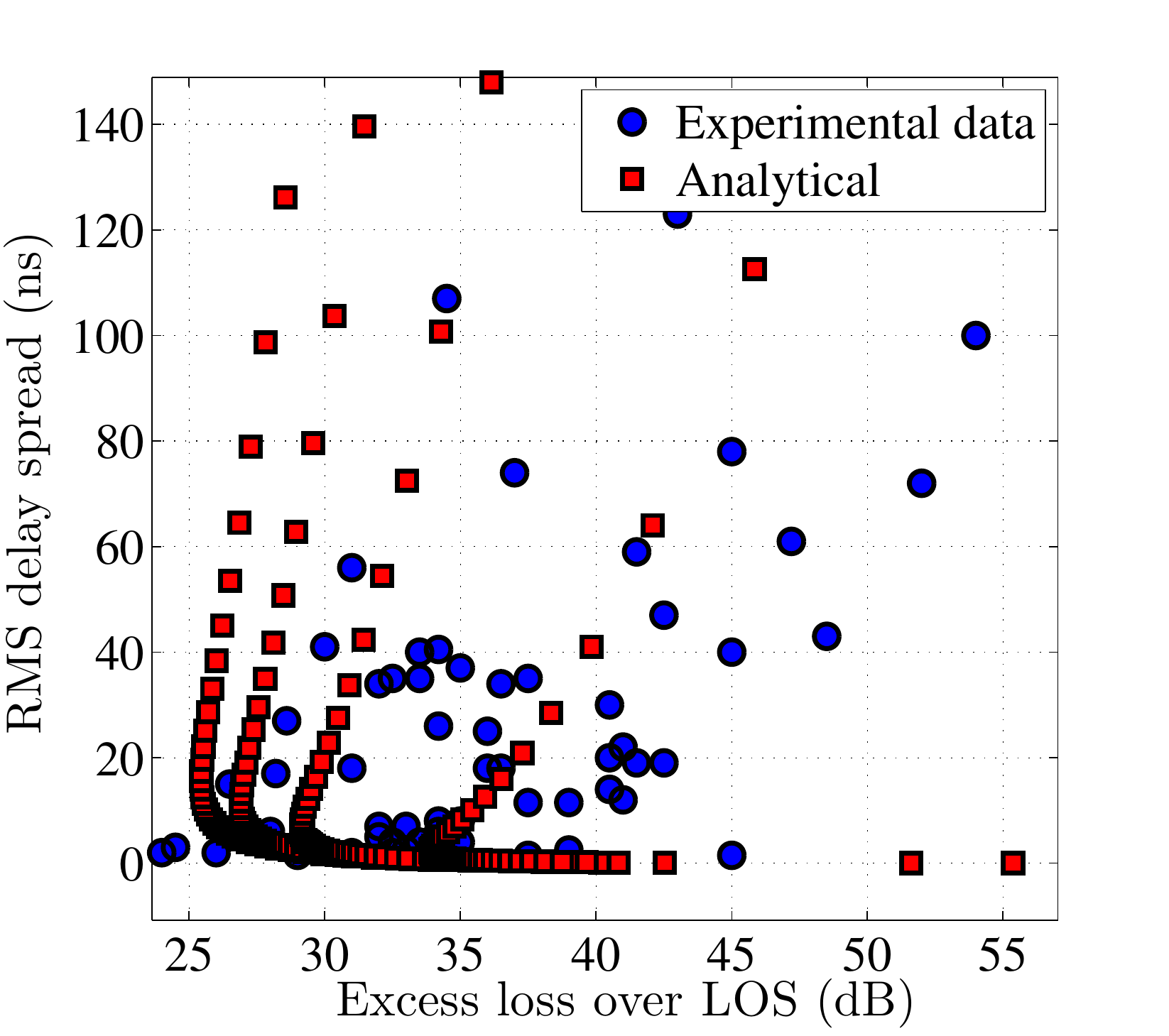}
 		\vspace{-0.65cm}
 		\caption{}
 	\end{subfigure}%
 	\caption{(a) RMS delay spread for varying antenna pointing direction, with $\lambda_{b}=8.75\times10^{-5}$, $\theta_{b}=7^{0}$, and $d=75$ m, (b) RMS delay spread for varying excess loss over LOS with reflections, with $\lambda_{b}=8.75\times10^{-5}$, $\theta_{b}=7^{0}$, and $d=75$ m.}
 	\label{fig:rms_del_spread_compare}
 \end{figure}
\begin{figure*}
	\centering	
	\includegraphics[width=0.24\textwidth]{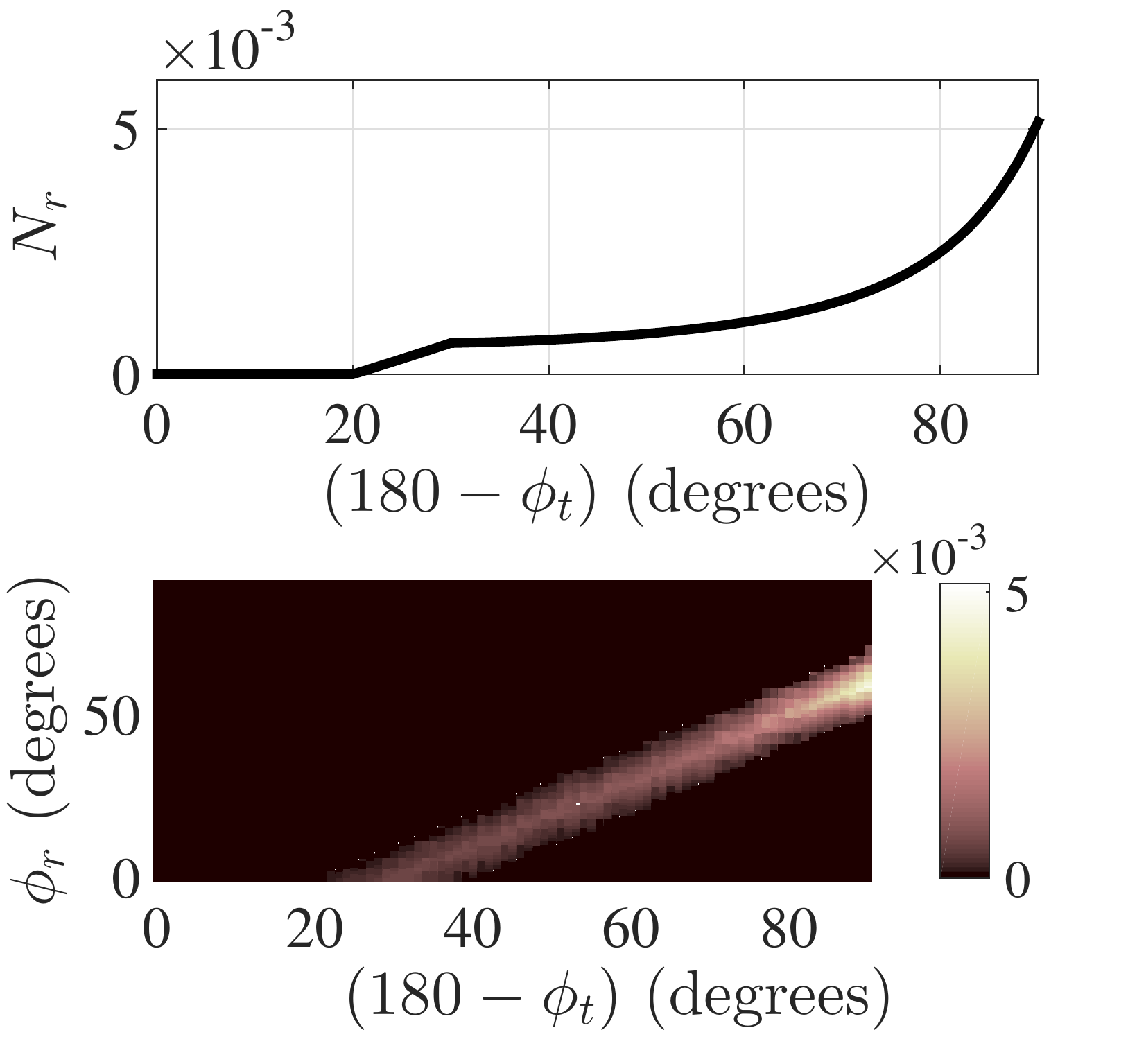}\hspace{-0.5em}
	\includegraphics[width=0.24\textwidth]{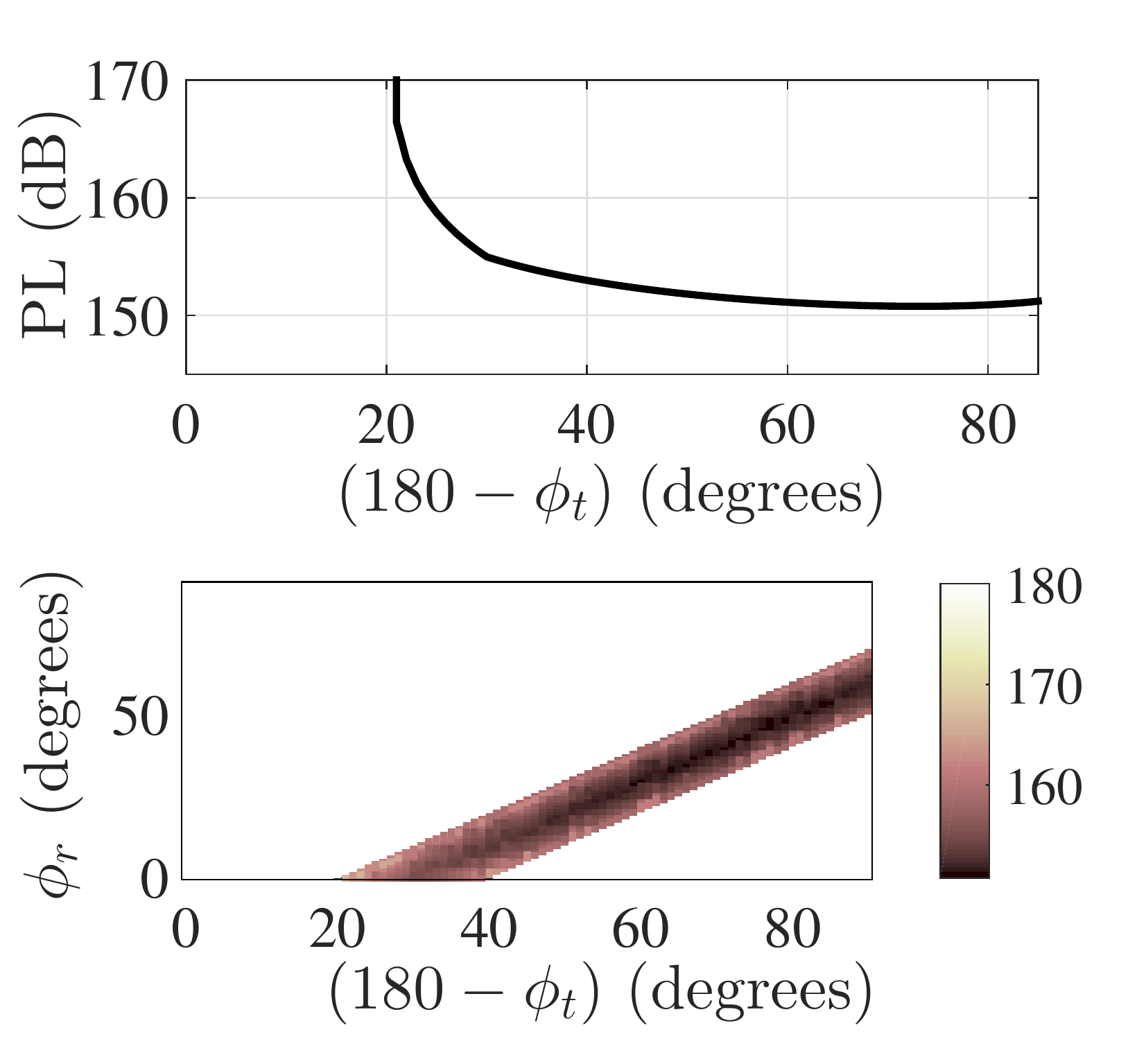}\hspace{-0.5em}
	\includegraphics[width=0.24\textwidth]{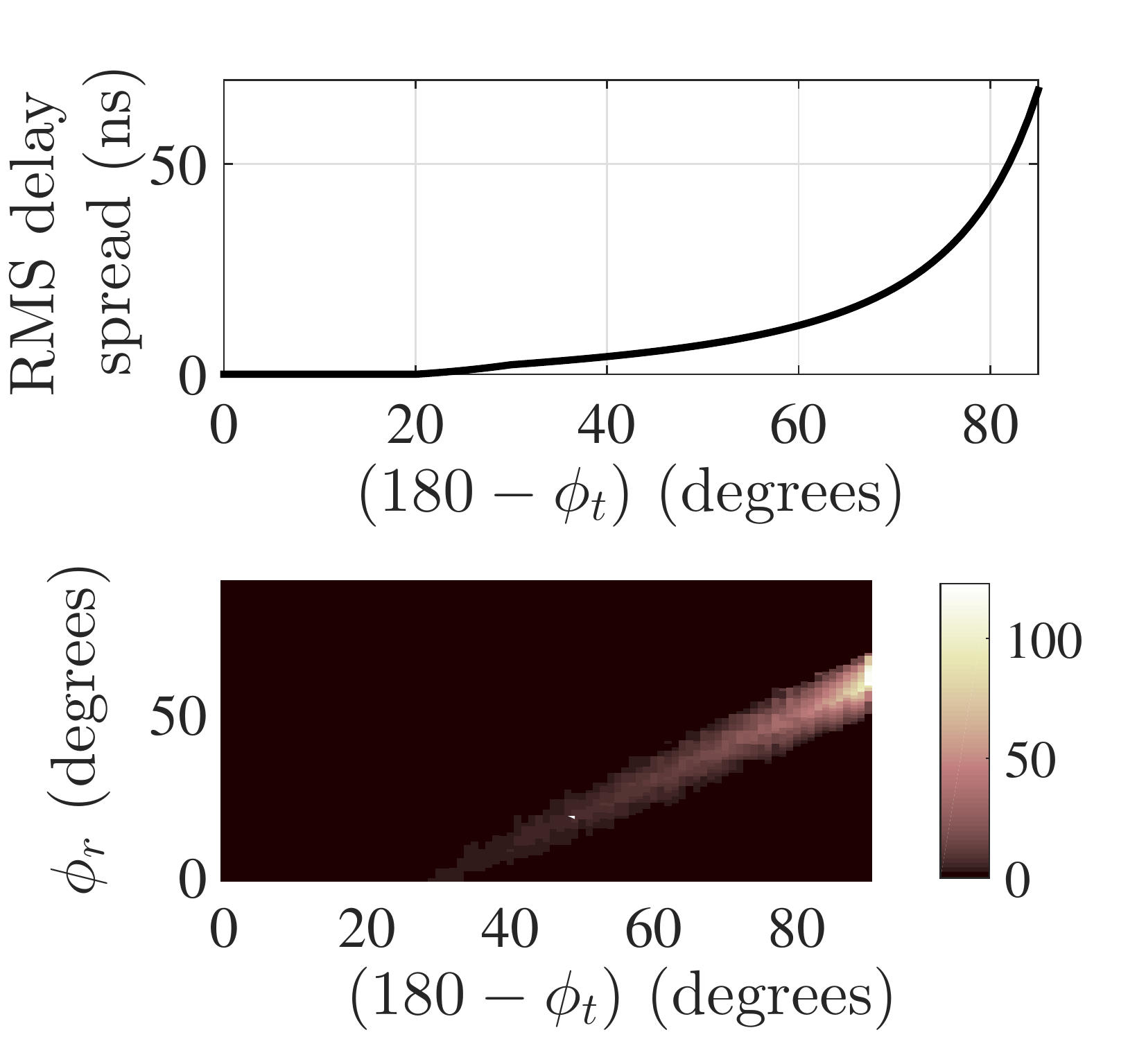}\hspace{-0.5em}
	\includegraphics[width=0.24\textwidth]{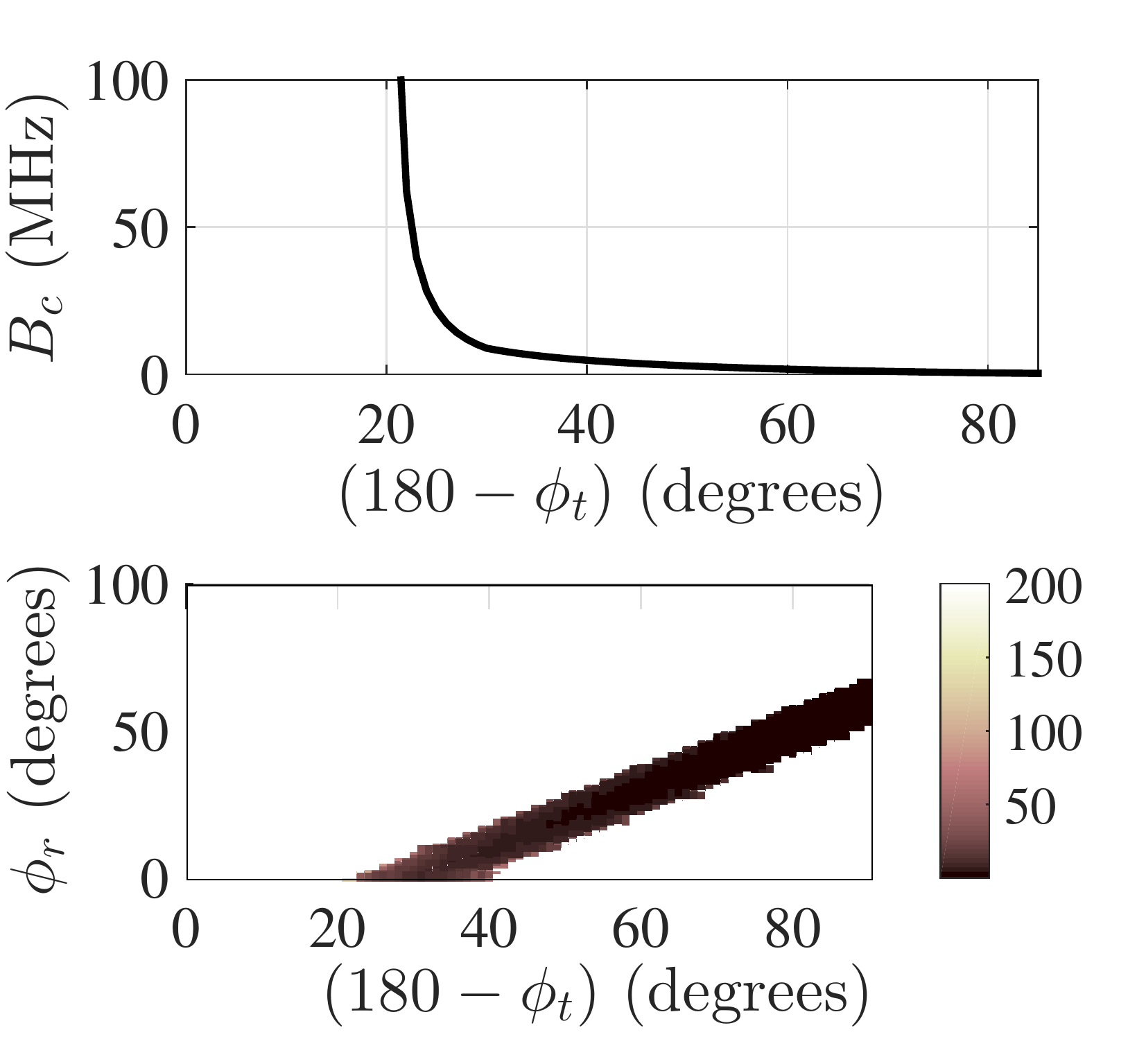}
	{$\mspace{1mu}$(a)$\mspace{210mu}$(b)$\mspace{210mu}$(c)$\mspace{210mu}$(d)}   
	\caption{Comparison of exact and approximated analytical expressions for $\lambda_{b}=12\times 10^{-5}$, $\E[l]=\E[w]=25$ m, $\lambda_{h}=20\times 10^{-4}$, $W_{h}=30$ cm, $P_{self}=0.25$, $d=50$ m, $\phi_{b}=15^{0}$, and $\theta_{b}=10^{0}$; (a) average number of first order reflection components vs sum of pointing angles of  transmitter-receiver antenna beams, (b) path loss vs sum of pointing angles of  transmitter-receiver antenna beams, (c) RMS delay spread vs sum of pointing angles of  transmitter-receiver antenna beams, (d) coherence bandwidth vs sum of pointing angles of  transmitter-receiver antenna beams.}
	\label{fig:performance_analysis_2}
	\vspace{-0.25cm}
\end{figure*}
\begin{figure*}
	\centering
	\includegraphics[width=0.24\textwidth]{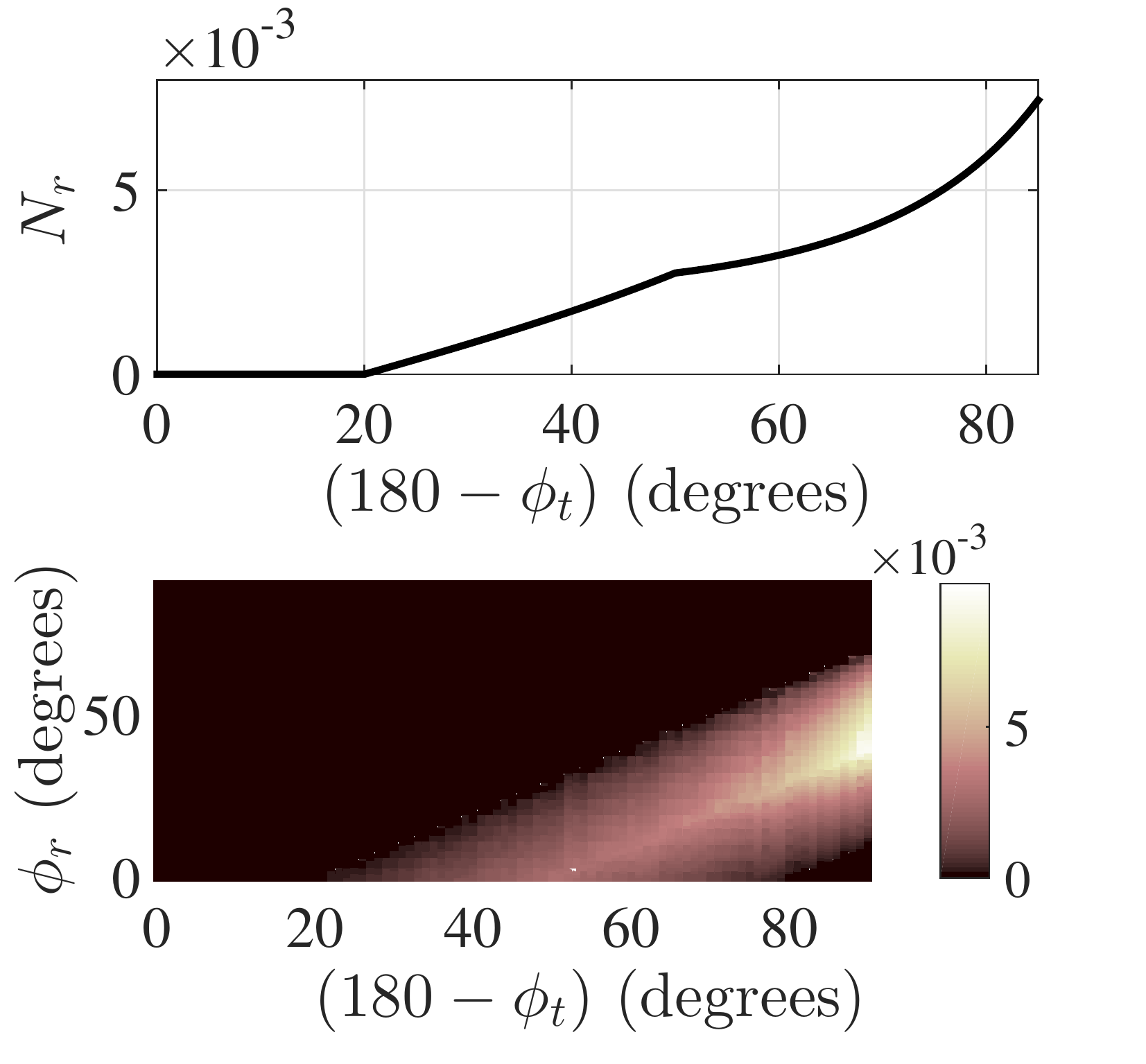}\hspace{-0.5em}
	\includegraphics[width=0.24\textwidth]{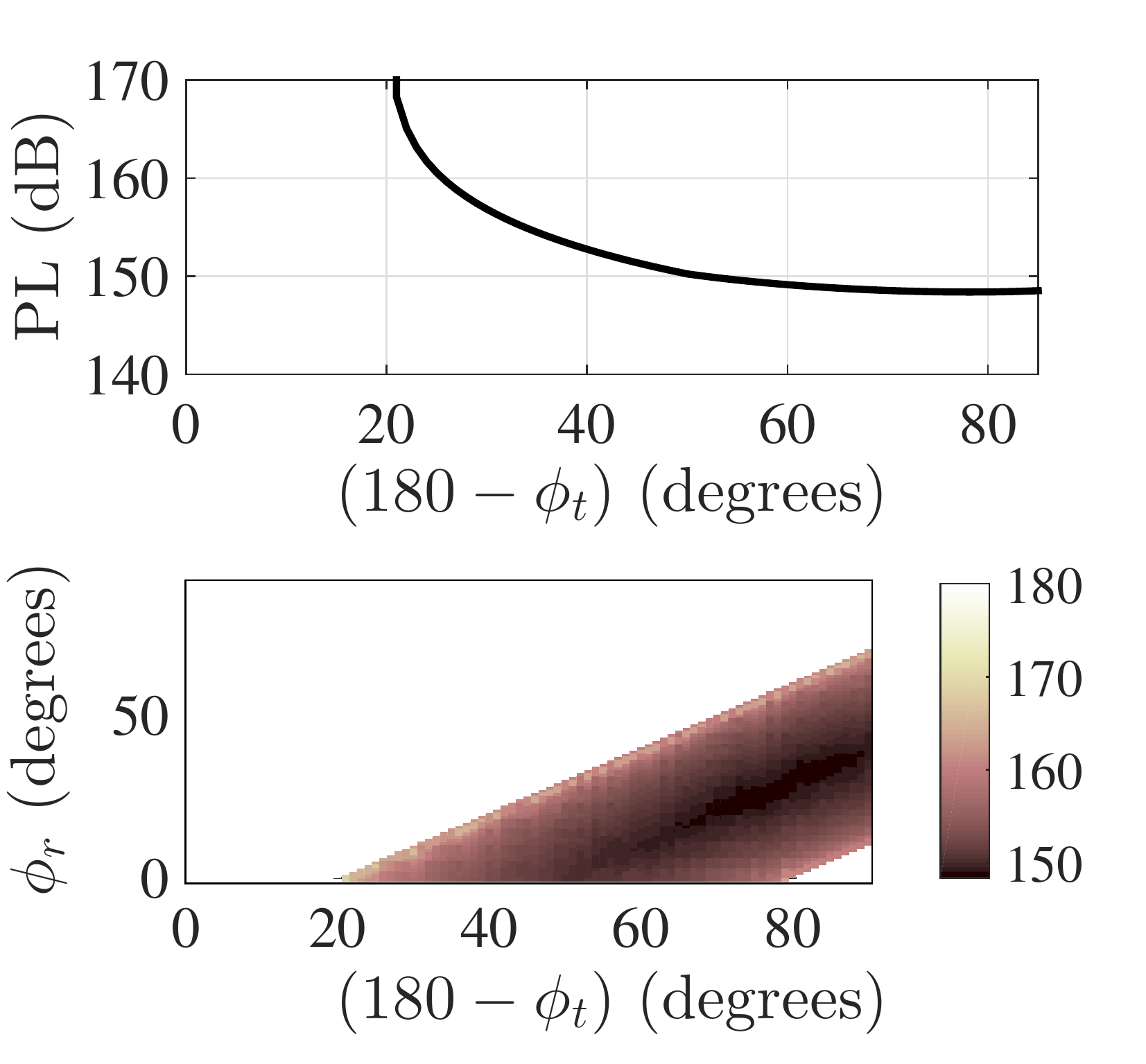}\hspace{-0.5em}
	\includegraphics[width=0.24\textwidth]{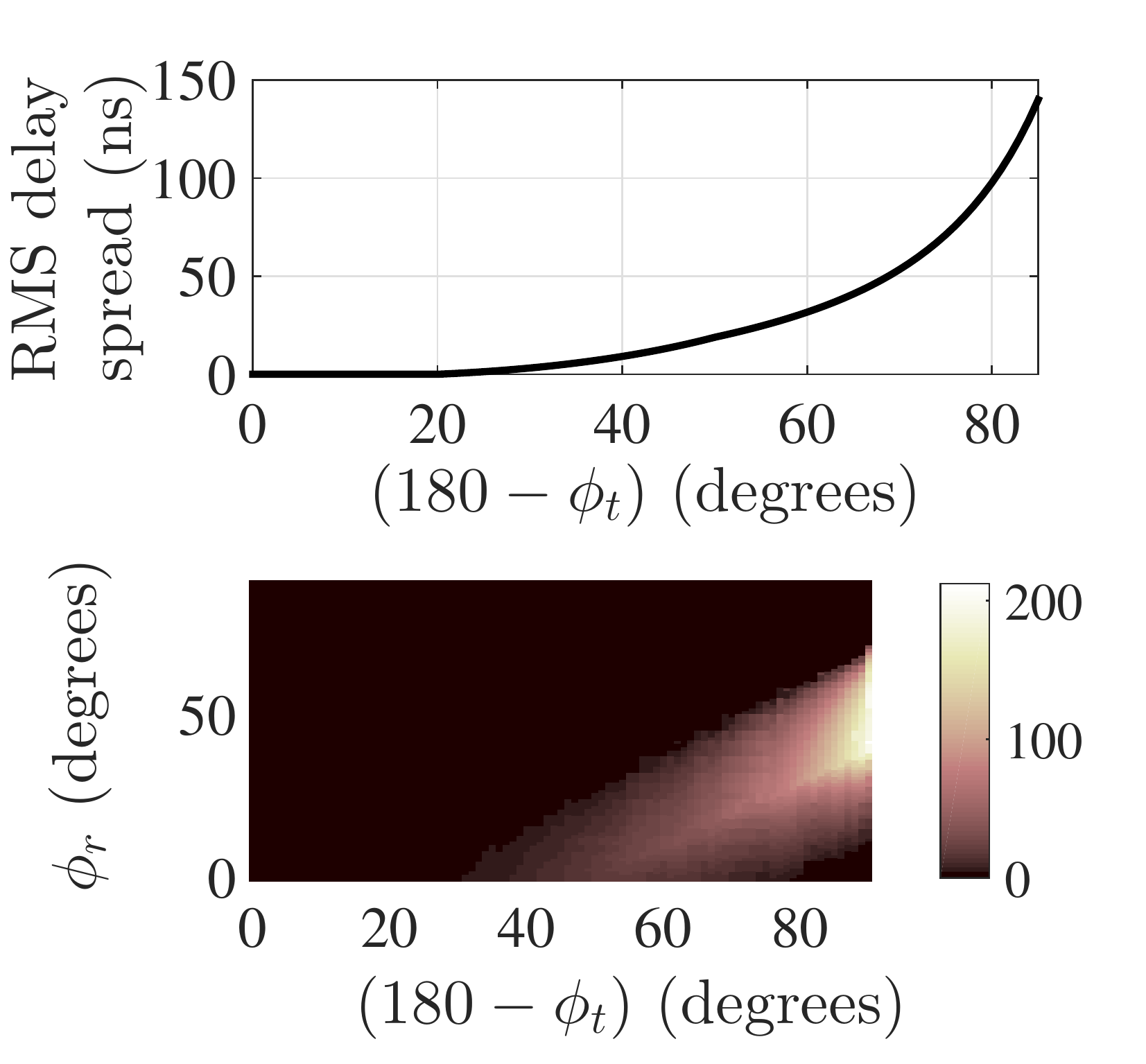}\hspace{-0.5em}
	\includegraphics[width=0.24\textwidth]{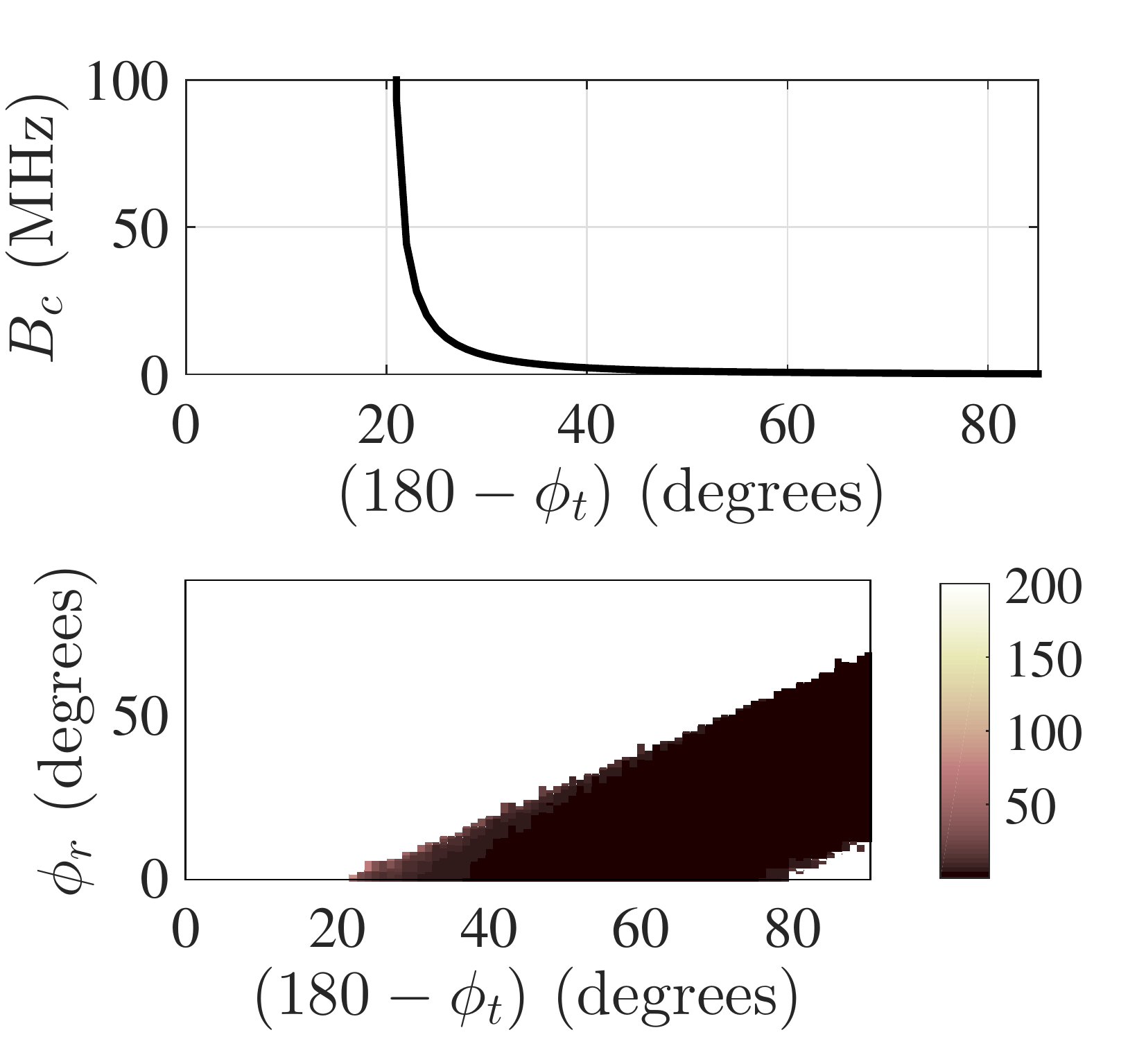}
	{$\mspace{1mu}$(a)$\mspace{210mu}$(b)$\mspace{210mu}$(c)$\mspace{210mu}$(d)}        
	\caption{Comparison of exact and approximated analytical expressions for $\lambda_{b}=12\times 10^{-5}$, $\E[l]=\E[w]=25$ m, $\lambda_{h}=20\times 10^{-4}$, $W_{h}=30$ cm, $P_{self}=0.25$, $d=75$ m, $\phi_{b}=25^{0}$, and $\theta_{b}=30^{0}$; (a) average number of first order reflection components vs sum of pointing angles of  transmitter-receiver antenna beams, (b) path loss vs sum of pointing angles of  transmitter-receiver antenna beams, (c) RMS delay spread vs sum of pointing angles of  transmitter-receiver antenna beams, (d) coherence bandwidth vs sum of pointing angles of  transmitter-receiver antenna beams.}
	\label{fig:performance_analysis_3}
	\vspace{-0.25cm}
\end{figure*} 

Finally, we compare the analytically evaluated RMS delay spread of directional NLOS mmWave channels (after excluding the term which models human blockage, i.e., $z=0$ in (\ref{rms_del})) with experimental results reported in \cite{rappaport201238} for peer-to-peer mmWave links in Fig.~\ref{fig:rms_del_spread_compare}(a) and Fig.~\ref{fig:rms_del_spread_compare}(b). The results are plotted for various pointing angles of transmitter and receiver antenna beams. To model the statistical parameters of buildings in The University of Texas at Austin campus, where the experiments were conducted \cite{rappaport201238}, we utilize Google map of the campus area and calculated, $\E[l]=55$ m and $\E[w]=54$ m, and an estimated building coverage of approximately 26\%. Based on the data, $\lambda_{b}$ is determined as, $\lambda_{b}=8.75\times 10^{-5}$.
Similar to the experimental setup, antenna HPBW $\theta_{b}=7^{0}$ and transmitter-receiver separation $d=75$ m (an average of the distance range from $19$ m to $129$ m considered in the experiment) are chosen for the analytical evaluation. Further, the building orientation is assumed to be $\phi_{b}=0^{0}$. The analytical plot in Fig.~\ref{fig:rms_del_spread_compare}(a) shows a trend consistent with experimental data, and yields a lower bound for RMS delay spread of the channel for most pointing angle combinations. This is due to the fact that the proposed analytical framework does not model any reflection components from lamp posts, vehicles etc., which is likely to increase RMS delay spread of the channel.
We also compare analytical and experimental values of RMS delay spread versus excess loss for NLOS signal propagation as compared to LOS propagation loss in Fig.~\ref{fig:rms_del_spread_compare}(b). Due to the unavailability of exact values for reflection coefficients of buildings at $38$ GHz operating frequency, we obtain corresponding values by linearly interpolating the value of reflection coefficients at $28$ GHz \cite{zhao201328} and $60$ GHz \cite{correia1994estimation} following the approach adopted in \cite{hur2016proposal}. We consider a concrete building surface ($\Gamma_{r,m}=3.18$ dB) for analysis, and plot RMS delay spread versus excess loss in Fig.~\ref{fig:rms_del_spread_compare}(b). Notably, the channel parameters evaluated based on the results depicted in Fig.~\ref{fig:performance_analysis} and Fig.~\ref{fig:rms_del_spread_compare} are consistent with experimental results reported in \cite{rappaport2015wideband}, \cite{rajagopal2012channel,rappaport2012cellular,rappaport201238}. In essence, the proposed analytical framework can be used for modeling  typical urban outdoor directional mmWave channels to a fair degree of accuracy.
\subsection{Analysis of the proposed channel model for varying channel environments} 
In this section, we first compare the approximated closed form expressions derived for average number of first order reflections $N_{r/\phi_{b}}$ and path loss for fixed orientation angle of $\phi_{b}$ for all buildings with corresponding exact expressions.
The exact values for $N_{r/\phi_{b}}$ and path loss are numerically calculated using the expressions, $N_{r/\phi_{b}}=\lambda_{b}\E_{L,W}\left[\int_{A}(1-P_{b/\phi_{b},\theta_{n}})\text{d}A\right]$ and $\frac{1}{PL_{avg/\phi_{b}}}=\Gamma_{l}\Gamma_{r}\lambda_{b}
 \int_{A}\frac{\text{sin}\theta_{n}\left(1-P_{b/\phi_{b},\theta_{n}}\right)}{\left(d\left|\text{cos}\phi_{b}\right|\text{sec}\theta_{n}\right)^{2}}\text{d}A$, respectively, with  $P_{b/\phi_{b},\theta_{n}}=1-(P_{self})^{i}\exp(-\lambda_{b}(d\left|\text{cos}\phi_{b}\right|(
 \E[l]\text{tan}\theta_{n}+\E[w])+\E\left[l\right]\E\left[w\right])-\lambda_{h}W_{h}d|\text{cos}\phi_{b}|\text{sec}\theta_{n})$. The corresponding approximated values are found out from (\ref{N_r3}) and  (\ref{avg_sig_pow3}), respectively. The parameters used for the comparison are: $\lambda_{b}=12\times 10^{-5}$, $\E[l]=\E[w]=25$ m, $\lambda_{h}=20\times 10^{-4}$, $W_{h}=30$ cm, $P_{self}=0.25$, $\phi_{b}=15^{0}$. As a case study, we assume that both nodes are carried by persons, and therefore, $i=2$. Average errors of $0.25$\% and $0.081$\% are observed for the exact and approximated expressions of $N_{r/\phi_{b}}$ and path loss, respectively, at $d=75$ m and $\theta_{b}=30^{0}$. It should be noted that maximum error for $N_{r/\phi_{b}}$ and path loss are approximately $28$\% and $35$\%, respectively at $\phi_{r}=40^{0}$ and $\phi_{t}=95^{0}$. However, average error for $N_{r/\phi_{b}}$ and path loss at $d=50$ m and $\theta_{b}=10^{0}$ are reduced to $0.031$\% and $6.27\times 10^{-3}$\%, respectively with the maximum error observed to be within $4$\% at $\phi_{r}=40^{0}$ and $\phi_{t}=95^{0}$. We also present $N_{r/\phi_{b}}$, path loss, RMS delay spread, and coherence bandwidth of directional mmWave channel corresponding to $d=50$ m, $\theta_{b}=10^{0}$, and $\phi_{b}=15^{0}$ using exact expressions with respect to various combinations of $\phi_{r}$ and $\phi_{t}$ in Fig.~\ref{fig:performance_analysis_2}. The coherence bandwidth $B_{c}$ is calculated with frequency correlation above $0.9$ (using the expression $B_{c}\approx\frac{1}{50\sigma_{\tau}}$). We plot the maximum or minimum value\footnote[5]{We plot maximum values of $N_{r/\phi_{b}}$ and minimum values of path loss for a given $180^{0}-\phi_{t}$. Similarly, maximum values of RMS delay spread and minimum values of $B_{c}$ are plotted with respect to $180^{0}-\phi_{t}$.} of a given channel parameter with respect to $\phi_{t}$ in each sub-figure of Fig.~\ref{fig:performance_analysis_2} (corresponding to the upper part of each sub-figure). The plots in Fig.~\ref{fig:performance_analysis_2}(a) reveal that large antenna beam pointing angles at the transmitter or receiver (or both) yield higher number of first order reflections as compared to small antenna beam pointing angles. However, as depicted in Fig.~\ref{fig:performance_analysis_2}(b), the path loss initially reduces and then increases with larger antenna beam pointing angle. This is due to the fact that signal power decreases due to higher blockage probability of reflection components and propagation distance with increase in antenna beam pointing angle. Initially, path loss is reduced due to the increase in number of reflection components; subsequently, however, it begins to increase since the effect of propagation losses and blockage dominates the rise in number of reflection components. The plots in Fig.~\ref{fig:performance_analysis_2}(c) show that RMS delay spread of channel increases sharply with respect to pointing angle. The results depicted in Fig.~\ref{fig:performance_analysis_2}(d) show that the coherence bandwidth sharply reduces from approximately $100$ MHz to $1$ MHz as $\phi_{r}$  increases from  $0^{0}$ to $60^{0}$. The result assumes relevance for the design of multi-carrier communication systems where coherence bandwidth of the channel eventually dictates the minimum number of sub-carriers and achievable throughput of the system. Thus, we see that the antenna pointing direction plays an instrumental role in reducing the complexity of a communication system in terms of number of sub-carriers and the ensuing need for channel equalizers with a higher number of taps (a fact also noted in \cite{rappaport201238}). The trend of the plot in Fig.~\ref{fig:performance_analysis_3} is similar to that observed in Fig.~\ref{fig:performance_analysis_2}, which is shown for $d=75$ m, $\theta_{b}=30^{0}$, and $\phi_{r}=25^{0}$. Key observations from Fig.~\ref{fig:performance_analysis_3} include: increased  number of reflection components, decreased path loss, increased RMS delay spread, and reduced coherence bandwidth as compared to the results depicted in Fig.~\ref{fig:performance_analysis_2}, which are along expected lines for a wireless link operating with wider antenna beams and larger transmitter-receiver separation.

We can also analytically evaluate the probability of occurence of the number of first order reflections, Prob($N^{i}$), from a Poisson probability mass function with the average number of reflections being numerically evaluated from the exact expression used for generating Fig.~\ref{fig:performance_analysis_2} and Fig.~\ref{fig:performance_analysis_3}. We compare the simulated and analytically evaluated values of Prob($N^{i}$) using Kullback-Leibler divergence (KLD) by considering blockage due to buildings only.  Due to the high complexity involved in finding probability using simulation in case of a large number of reflection components, we limit maximum value of $N^{i}$ as $2$.
The parameters chosen for the comparison are: $\lambda_{b}=12\times 10^{-5}$, $\E[l]=\E[w]=25$ m, $\phi_{b}=15^{0}$, $\phi_{r}=51^{0}$, $\phi_{t}=94^{0}$ (selected based on Fig.~\ref{fig:performance_analysis_2}(a)), and $\theta_{b}=30^{0}$.
We consider two scenarios with $d=75$ m and $150$ m, and calculate KLD equal to $-0.0133$ and $2.669\times 10^{-4}$, respectively. The results confirm that the Poisson assumption for number of reflection component is valid for large transmitter receiver separation distances.
\section{Conclusion}
In this paper, we have proposed a novel geometry based model to evaluate average number of first order reflection components, path loss, average multi-path delay, and root mean square delay spread of mmWave NLOS directional channels. The joint effect of building density, antenna half power beamwidth, antenna beam pointing direction, and transmission distance on NLOS channel characteristics is investigated.
The results show that the scenario with larger transmission distance and broad antenna pointing direction possesses much more multi-path components and higher root mean square delay spread as concluded by the available experimental results. In addition, path loss reduces with initial increase in antenna beam pointing direction and begins to rise sharply after a critical point.
Moreover, the findings based on the numerical evaluation fairly agree with the experimental data reported for outdoor NLOS peer-to-peer mmWave links with operating frequency of $38$ GHz. Therefore, investigations of this paper shed light into mmWave directional NLOS channel characteristics and provide new avenues for the future research on mmWave network design, and its performance study.

As a future study, it would be interesting to develop a three dimensional analytical channel model for mmWave outdoor NLOS communications. Another direction could be the application of the proposed channel model for the performance study of outdoor mmWave networks.
\begin{appendices}
\section{Derivation of the expression for elemental area}
The elemental area $\text{d}A$ can be derived from the expression of $A$ in (\ref{prob_coup}) as,\\
\setcounter{equation}{30}
\begin{align}\label{ele_area}
\text{d}A&=\text{d}a\frac{d\left|\text{cos}\phi_{b}\right|}{2}\left(\text{tan}\left(\theta_{n}+\text{d}\theta_{n}\right)
-\text{tan}\theta_{n}
\right)\nonumber \\
&=\frac{d\left|\text{cos}\phi_{b}\right|}{2}\text{sec}^{2}\theta_{n}\left(\frac{\text{d}\theta_{n}}{1-\text{d}\theta_{n}\text{tan}
	\theta_{n}}\right)\text{d}a, 
\end{align}
where $\text{d}a$ and $\text{d}\theta_{n}$ are the incremental values for $a$ and $\theta_{n}$, respectively. In fact, $d\theta_{n}\text{tan}\theta_{n}\approx 0$ for smaller value of $\text{d}\theta_{n}$ and $\theta_{n}<\pi/2$ (the condition $\theta_{n}=\pi/2$ corresponds to an infinite propagation distance (see Fig.~\ref{fig:Analytic1}(a)) which does not have any impact on the received power), (\ref{ele_area}) is simplified as,\\
\begin{align}\label{el_area}
\text{d}A=\frac{d\left|\text{cos}\phi_{b}\right|}{2}\text{sec}^{2}\theta_{n}\text{d}\theta_{n}\text{d}a.	
\end{align}
\section{Derivation of PDFs $f_{\hat{\Theta}_{tl}^{f}}\left(\hat{\theta}_{tl}^{f}/\hat{d}\right)$ and $f_{\hat{D}}(\hat{d})$}
Let $A'_{\overline{d}}$ denote the area of feasible region for first reflection with $d_{max}=\overline{d}$. Therefore, cumulative distribution function (CDF) for $\hat{d}$ is determined as,\\
\begin{align}
\hspace{-0.2cm}F_{\hat{D}}(\hat{d})=\frac{A'_{\hat{d}}}{A'}=\frac{C_{1}\left(\hat{d}-\frac{a'}{2}\right)+C_{2}\left(\hat{d}^{2}-\left(\frac{a'}{2}\right)^{2}\right)}{C_{1}\left(d_{max}-\frac{a'}{2}\right)+C_{2}\left(d_{max}^{2}-\left(\frac{a'}{2}\right)^{2}\right)}
\end{align}
where $C_{1}=\frac{a}{2}\left( \frac{\text{sin}\theta_{tu}^{f}}{\text{cot}\theta_{tu}^{f}}+\frac{\text{sin}\theta_{ti}^{f}}{\text{cot}\theta_{ti}^{f}}\right)$, $C_{2}=\frac{\text{cot}\theta_{tu}^{f}-\text{cot}\theta_{ti}^{f}}{2}$. Therefore, PDF for $\hat{d}$ is obtained as,\\
\begin{align}
f_{\hat{D}}(\hat{d})=\frac{C_{1}+2C_{2}\hat{d}}{C_{1}\left(d_{max}-\frac{a'}{2}\right)+C_{2}\left(d_{max}^{2}-\left(\frac{a'}{2}\right)^{2}\right)}.
\end{align}
The conditional PDF of $\hat{\theta}_{ti}^{f}$ $\left(f_{\hat{\Theta}_{ti}^{f}}\left(\hat{\theta}_{ti}^{f}/\hat{d}\right)\right)$ is derived based on the illustrations provided in Fig.~\ref{fig:second_order_reflection} and Fig.~\ref{fig:feasible_area_second_order_reflection}. We note that the location of the points in a PPP is uniformly distributed for a given number of points in an euclidean space \cite{haenggi2012stochastic}.
Let $h+\frac{a}{2}$ represents the vertical distance between the center of a reflector and the line passing through the points $Tx$ and $Tx_{im}^{f1}$. Based on the definition $\hat{\theta}_{ti}^{f}=\text{tan}^{-1}\frac{h}{\hat{d}-a'/2}=g(h)$, where the distance $h$ is uniformly distributed with lower and upper limits $h_{l}$ and $h_{u}$, respectively, the conditional PDF of $\hat{\theta}_{ti}^{f}$ is obtained as,\\
\begin{align}\label{PDF_theta_n}
f_{\hat{\Theta}_{ti}^{f}}(\hat{\theta}_{ti}^{f}/\hat{d})=&f_{H}(g^{-1}(\theta_{n}))\left|\frac{d((\hat{d}-a'/2)\text{tan}\hat{\theta}_{ti}^{f})}{d\hat{\theta}_{ti}^{f}}\right|\nonumber\\=& \left\{\hspace{-0.2cm}\begin{array}{lr}
\frac{\hat{d}-a'/2}{h_{u}-h_{l}}\text{sec}^{2}\hat{\theta}_{ti}^{f}, & \theta_{l}\leq\hat{\theta}_{ti}^{f}\leq\theta_{u}\\\\
0, & \text{otherwise.}
\end{array}\right.
\end{align}
It is observed from Fig.~\ref{fig:second_order_reflection} and Fig.~\ref{fig:feasible_area_second_order_reflection} that $h_{u}$ and $h_{l}$ are decided by the location of the center of a reflector with respect to the feasible region $A'$. Therefore, first reflection is possible to originate only if $h_{u}=(\hat{d}-a'/2)\text{tan}(\frac{\pi}{2}-\theta_{tu}^{f})$ and $h_{l}=(\hat{d}-a'/2)\text{tan}\theta_{li}^{f}$, where $\theta_{li}^{f}=\text{tan}^{-1}\left(\text{cot}\theta_{ti}^{f}-\frac{a}{(\hat{d}-a'/2)}\right)$. Hence (\ref{PDF_theta_n}) is modified into,\\
\begin{align}\label{PDF_theta_n_int}
f_{\hat{\Theta}_{ti}^{f}}(\hat{\theta}_{ti}^{f}/\hat{d})=\left\{\hspace{-0.2cm}\begin{array}{lr}
\frac{\text{sec}^{2}\hat{\theta}_{ti}^{f}}{\text{tan}(\frac{\pi}{2}-\theta_{tu}^{f})-\text{tan}\theta_{li}^{f}}, & \theta_{li}^{f}\leq\hat{\theta}_{ti}^{f}\leq\frac{\pi}{2}-\theta_{tu}^{f}\\\\
0, & \text{otherwise.}
\end{array}\right.
\end{align}
It can also inferred that $\hat{\theta}_{ti}^{f}=\frac{\pi}{2}-\theta_{ti}^{f}$ for $\hat{\theta}_{ti}^{f}<\theta_{ti}^{f}$. Therefore, (\ref{PDF_theta_n_int}) is simplified as (\ref{PDF_theta_n_final}).
\end{appendices}

\end{document}